\documentclass[final]{jfm}

\usepackage{graphicx}
\usepackage{newtxtext}
\usepackage{newtxmath}
\usepackage{natbib}
\usepackage{hyperref}
\usepackage{placeins}
\usepackage[singlelinecheck=false, justification=justified, width=\textwidth]{caption}

\hypersetup{
    colorlinks = true,
    urlcolor   = blue,
    citecolor  = black,
}

\newcommand{\RomanNumeralCaps}[1]
\linenumbers

\DeclareMathOperator\erf{erf}
\DeclareMathOperator\erfi{erfi}

\title{Unsteady aerodynamic response of pitching airfoils represented by Gaussian body forces}

\author{Emanuel Taschner \aff{1}
  \corresp{\email{e.taschner@tudelft.nl}},
  Georgios Deskos \aff{2},
  Michael B Kuhn \aff{2},
  Jan-Willem van Wingerden \aff{1}
 \and Luis A Mart{\'i}nez-Tossas\aff{2}}

\affiliation{\aff{1} Delft University of Technology, Delft, South Holland, The Netherlands
\aff{2}National Renewable Energy Laboratory, Golden, CO, USA}

\begin{document}
\maketitle

\begin{abstract}
The actuator line method (ALM) is an approach commonly used to represent lifting and dragging devices like wings and blades in large-eddy simulations (LES).
The crux of the ALM is the projection of the actuator point forces onto the LES grid by means of a Gaussian regularisation kernel. The minimum width of the kernel is constrained by the grid size; however, for most practical applications like LES of wind turbines, this value is an order of magnitude larger than the optimal value which maximises accuracy. This discrepancy motivated the development of corrections for the actuator line, which, however, neglect the effect of unsteady spanwise shed vorticity.
In this work, we develop a model for the impact of spanwise shed vorticity on the unsteady loading of an airfoil modelled as a Gaussian body force. The model solution is derived both in the time and frequency domain and features an explicit dependence on the Gaussian kernel width. 
We validate the model with LES within the linear regime of the lift curve for both pitch steps and periodic pitching with reduced frequencies of $k=0.1, 0.2 \text{ and } 0.3$. The Gaussian kernel width affects, in particular, the amplitude of the unsteady lift, which can be up to 40\% smaller than the quasi-steady amplitude within the considered range of reduced frequencies and kernel widths.
\end{abstract}


\section{Introduction}
\label{sec:Intro}
Today's wind energy generation relies on large-scale wind farms operating optimally in the lower (surface) part of the  atmospheric boundary layer (ABL) \citep{veers_grand_2019, porte-agel_wind-turbine_2020}. Optimal performance of wind farms requires accurate prediction of the aerodynamic characteristics of the individual turbines and their interaction with wakes and the ambient atmospheric flow. The unsteady ABL flow is commonly computed using large-eddy simulation (LES), where wind turbines are parameterised with actuator models like actuator disks or actuator lines \citep{breton_survey_2017}. \par
The actuator line model (ALM), initially developed by \citet{sorensen_numerical_2002}, represents lifting and dragging devices, such as blades and wings, as discrete lines. It allows for the distribution of body forces along these lines, effectively simulating the influence of solid bodies on the flow. As such, when  the ALM is utilised to model a wind turbine rotor, it is capable of capturing the footprint of the individual blades and the associated vortex system composed of the bound, tip and root vortices \citep{troldborg_actuator_2009, troldborg_numerical_2010, ivanell_stability_2010, martinez-tossas_large_2015}. Nevertheless, the ALM is still relatively simple to implement thanks to its suitability with Cartesian grids. This balance between simplicity, physical accuracy, and computational cost has contributed to the ALM's ongoing popularity more than twenty years after its inception \citep{sanderse_review_2011, breton_survey_2017}. During these years, the ALM has enabled many advancements in wind energy and beyond, for example, the study of ABL-turbine interaction \citep{troldborg_numerical_2011, churchfield_numerical_2012}, the analysis of near-wake vortex dynamics like the tip vortex instability \citep{sarmast_mutual_2014, sorensen_simulation_2015, kleusberg_tipvortex_2019,hodgkin_numerical_2022}, the study of novel wake control strategies \citep{yilmaz_optimal_2018, frederik_helix_2020}, or the simulation of vertical axis wind turbines \citep{mendoza_near-wake_2019} and complex wing configurations \citep{kleine_simulating_2023}.

\subsection{Gaussian force regularisation and smearing corrections for the ALM}
The accuracy of the ALM approach depends on the force projection from the discrete lines onto the LES grid using a regularisation kernel, $\eta_\epsilon$. In its most general form in three-dimensional space, the regularisation of a single discrete actuator point force $\boldsymbol{f}^{Act}$ across the volume $\Omega$ takes the following form,
\begin{equation}
    \boldsymbol{f}^{Reg}=\iiint_\Omega \boldsymbol{f}^{Act} \eta_\epsilon \,\mathrm{d} \boldsymbol{x}, \, \boldsymbol{x}\in\Omega \,\,\,  \text{with} \,\,\, \eta_\epsilon = \frac{1}{\pi^{3/2}\epsilon^3} e^{-(|\boldsymbol{x}-\boldsymbol{x}^{Act}|)^2/\epsilon^2},
\end{equation}
where $\epsilon$ is the Gaussian kernel width and $\boldsymbol{x}^{Act}$ the actuator point location. 
The need for calculating a convolution of the actuator forces with a regularisation kernel was introduced in the foundational ALM work by \citet{sorensen_numerical_2002} in order to avoid singular behaviour which would result from direct application of the actuator point forces on the LES grid. In their work, the shape of the regularisation kernel was chosen to be an isotropic Gaussian as introduced above, where the kernel width is constant along the entire span of the blade. 
Motivated by the idea that the actuator forces are ultimately supposed to model the dimension and shape of the wing or blade and the associated airfoils, a comprehensive body of literature exists aiming to improve the force regularisation accordingly \citep{martineztossas_optimal_2017, churchfield_advanced_2017, jha_actuator_2018, caprace_lifting_2019, schollenberger_boundary_2020, liu_evaluating_2022}.
However, independent of the employed kernel type and shape, the minimal allowable kernel width is dictated by the grid size, $\Delta x$, employed in LES, and arguments of a minimum allowable ratio,  $\epsilon/\Delta x$, have been made in order to avoid numerical oscillations and obtain convergence. Its proposed value ranges from $\epsilon/\Delta x=2$ \citep{troldborg_numerical_2010}  to $\epsilon/\Delta x \geq 4$ \citep{shives_mesh_2013} and $\epsilon/\Delta x \geq 5$ \citep{martinez-tossas_large_2015}. The resulting kernel widths from these considerations are often an order of magnitude larger than the optimal kernel width,~\footnote{An optimal kernel maximises the agreement between the flow solutions around an airfoil with chord, $c$, modelled as Gaussian body force akin to the ALM, and the corresponding potential flow solution.} which was found to be in the order of $\epsilon_{Opt}/c=\mathcal{O}(10^{-1})$ by \citet{martineztossas_optimal_2017} in agreement with the earlier proposed range of $\epsilon_{Opt}/c \in [1/8, 1/4]$ \citep{shives_mesh_2013}.\par
In practice, the existing discrepancy between the conditions for the optimal and numerically allowable kernel width is the reason why the standard ALM, when employed on coarse LES grids, fails to accurately reproduce the loading in the proximity of the wing and blade tips. One of the earliest proposals to address this issue was made by \citet{shen_tip_2005-1} who introduced a tip loss correction for the ALM similar to the ones employed for blade element momentum theory. Progress has been made on this issue by \citet{dag_new_2020} (originally \citet{dag_combined_2017}), who observed that the bound vortex created by the ALM is also of Gaussian shape and thus similar to the shape of a Lamb-Oseen vortex. This observation was shown to be a mathematical consequence of the Gaussian force regularisation by \cite{forsythe_coupled_2015} for the bound vortex and by \cite{martinez-tossas_filtered_2019} for the vorticity shed by an actuator line.\par
These insights enable the correction of ALM results based on the mismatch between the induced velocities along the actuator line as they result from shed Lamb-Oseen vortices, with the kernel width dictated by numerical considerations and from a reference considered optimal \citep{meyer_forsting_vortex-based_2019, martinez-tossas_filtered_2019, dag_new_2020, kleine_non-iterative_2023}. These corrections---also often labelled as smearing corrections \nolinebreak---were applied to wind turbine rotors, for example, by \citet{meyer_forsting_wake_2019}, \citet{stanly_large-eddy_2022} and \citet{taschner_new_2024}.\par
\citet{meyer_forsting_vortex-based_2019} and \cite{kleine_non-iterative_2023} also tested their corrections for unsteady operating conditions, namely the NREL 5MW turbine subject to a step in blade pitch or operating in sheared inflow. In these conditions, spanwise vorticity would also be shed in response to the varying strength of the bound vortex. The effect of this additional unsteady vorticity component is not captured by the previously mentioned corrections, which also neglect the influence of drag. \cite{kleine_non-iterative_2023} pointed out in their work that a better understanding of drag and unsteady effects and their relation to the error caused by Gaussian regularisation with large kernel widths could help to further decrease associated errors. The impact of drag was already studied in the work of \cite{caprace_immersed_2020}  who developed an immersed lifting and dragging line method, which captures streamwise/spanwise and normal/spanwise shed vorticity for lift and drag forcing, respectively.

\subsection{Unsteady aerodynamics in the context of the ALM}
The previous outline shows that progress has been made on developing the ALM, and theoretical insights lead to corrections for practically employed kernel widths. These developments mostly focused on steady and quasi-steady conditions. The operating conditions of a wind turbine are, however, subject to various periodic and aperiodic sources of unsteadiness like shear, veer and high-frequency turbulence fluctuations in the atmospheric inflow; body motion due to flexible structures; tower shadow effects; and interaction of the blades with wakes generated by upstream turbines \citep{leishman_challenges_2002}. There also exists a variety of wind farm control strategies to mitigate  wake effects on downstream turbines. These wake control strategies utilise the turbine's yaw, rotational and pitch degrees of freedom, often introducing additional sources of unsteadiness \citep{meyers_wind_2022}. \par
The body force approach of the ALM is at the core of its simplicity, and therefore it may not resolve phenomena associated with boundary-layer dynamics, e.g., flow separation/reattachment, laminar-turbulent transition and dynamic stall effects. Their effect on the unsteady lift and drag can only be included by means of additional models, e.g., the dynamic stall model from \cite{leishman_semi-empirical_1989}.
However, even for flow scenarios corresponding to unsteady attached flow, the question remains of how Gaussian force regularisation with sub-optimally large kernel widths may impact the unsteady loading. Vorticity shed in response to the wake suffers from an excess of smearing, similar to the steady case, affecting its feedback to the angle of attack (induction) and thus the loading on blades. The role of shed unsteady spanwise vorticity has been so far neglected from previous ALM corrections \citep{meyer_forsting_vortex-based_2019, martinez-tossas_filtered_2019,  dag_new_2020, kleine_non-iterative_2023}, and therefore it is the focus of this study. \par
We begin by tackling the unsteady two-dimensional incompressible, inviscid flow over a Gaussian body force, similar to the successive theoretical advancements of the ALM in steady conditions \citep{martineztossas_optimal_2017}. This approach links the unsteady two-dimensional ALM (or actuator point rather than actuator line) to the wealth of foundational work on the unsteady incompressible, inviscid thin-airfoil problem \citep{von_karman_airfoil_1938, bisplinghoff_aeroelasticity_1996,leishman_principles_2002}.\par
Unsteady inviscid solutions can be traced back to the pioneering  work of \citet{prandtl_uber_1924}. They noted that unsteady airfoil loading implies the shedding of circulation into the wake and suggested a first-order solution to the problem of a flapping airfoil. \citet{birnbaum_ebene_1924} solved this problem by utilising a series approximation and quantified the degree of unsteadiness with the so-called reduced frequency, $k$, a non-dimensional number given by the ratio of the time scales of the unsteady phenomena and the time needed by the flow to travel across the airfoil's semi-chord length. Since a number of solutions have been derived to address the unsteady thin-airfoil problem, it is helpful to categorise them based on two criteria \citep{leishman_principles_2002}: firstly, whether the solution is derived in the time or frequency domain, and secondly, whether the source of unsteadiness is the inflow (sinusoidal gusts of the normal velocity) or the airfoil motion, e.g., translatory oscillation (heave) and/or rotational oscillation (pitching).
\citet{wagner_uber_1925} and \citet{kussner_zusammenfassender_1936} derived time domain solutions for the cases of unsteady body motion and unsteady inflow, respectively. Corresponding frequency domain solutions were derived by \citet{theodorsen_general_1935} for the case of airfoil motion and by \citet{sears_aspects_1941} for gusts.\par
In either case, the lift can be split into three contributions: (i) quasi-steady, (ii) apparent-mass and (iii) wake-induced \citep{von_karman_airfoil_1938, sears_aspects_1941}. The computation of contribution (iii) is based on the transfer functions $T(k)$ and $S(k)$ for Theodorsen's and Sears's solutions, respectively. These transfer functions incorporate the frequency-dependent phase and magnitude modulation of the unsteady lift caused by the shed vorticity in the wake. In principle, the ALM inherently captures the wake-induced unsteady lift since it is modelled by the LES. However, sub-optimally large kernel widths cause inaccurate unsteady lift predictions due to excessive smearing of the shed vorticity, where the error is expected to depend on both Gaussian kernel width and reduced frequency.

\subsection{Objective}
The main contribution of the present work is the development of analytical velocity and vorticity solutions for the unsteady attached flow over a pitching airfoil modelled as a two-dimensional Gaussian body force. Based on these solutions, a first-order model for the unsteady loading on the airfoil is derived. The model is akin to the aerodynamic models of \citet{wagner_uber_1925} and \citet{theodorsen_general_1935}, but stems from a formulation of the governing equations which is consistent with the ALM. The developed model quantifies the effect of the shed spanwise vorticity on the phase and magnitude of the unsteady airfoil loading with an explicit dependence on the kernel width. It thus shows how the ability of the ALM to capture unsteady aerodynamics is affected by the width of the Gaussian kernel and the time scale of the source of unsteadiness. Here, the considered source of unsteadiness is the oscillatory pitch motion of the airfoil. Nevertheless, the developed first-order model could also be adapted to heave motions or sinusoidal gusts.\par
The remainder of the paper is organised as follows. The formulation of the unsteady two-dimensional problem and its general solution in terms of vorticity is derived in Section \nolinebreak \ref{Sec:UnsteadyGaussianBodyForces}. In Section \ref{sec:VelocitySolution}, solutions for the induced velocity due to an unsteady loading are derived. Section \ref{sec:FeedbackProblemSolution} then builds upon these solutions to obtain a model for the unsteady airfoil loading both in time and frequency domain. In Section \ref{sec:AnalyticVortSol}, an analytical solution for the time-dependent vorticity field is derived. The model predictions for the airfoil's unsteady loading are then compared to LES in Section \nolinebreak \ref{Sec:LESvsTheory}. Furthermore, the impact of neglecting non-linear contributions for the derivation of the theoretical solutions is assessed in Section  \ref{Sec:NonLinError} utilising the derived vorticity solution and the vorticity field from LES. We conclude with Section \ref{Sec:Conclusions}, where the findings are summarised and an outlook of future work is provided.
\section{Flow over an unsteady two-dimensional Gaussian body force}
\label{Sec:UnsteadyGaussianBodyForces}
The overarching goal of the following analysis is to obtain the unsteady loading on an airfoil modelled as two-dimensional Gaussian body force both in terms of the phase and magnitude modulations observed with respect to the quasi-steady reference. Our approach comprises three main steps: First, a general form of the solution for the unsteady vorticity field around the airfoil is derived.  Second, by means of the Biot-Savart law the corresponding induced velocity solution at the airfoil location is obtained, which alters the angle of attack and thus the unsteady loading (Sections \ref{sec:VelocityNormSolution} and \ref{sec:VelocityStreamwiseSolution}). Third, the unsteady loading couples to the generation of shed vorticity resulting in a feedback problem which can be solved either in the time domain by formulating a root finding problem (Section \ref{sec:TimeDomainSolution}) or in the frequency domain by deriving the system's closed-loop transfer function (Section \ref{sec:FrequencyDomainSolution}). In this section we focus on the first step, i.e., the derivation of the general unsteady vorticity solution.\par 
For the derivations we assume incompressible flow in the infinite Reynolds number limit. Following the work of \cite{martineztossas_optimal_2017}, but retaining the time derivative term, our starting point is the two-dimensional unsteady Euler equation in non-dimensional form. We use the free stream velocity, $U_\infty$, and the airfoil's chord length, $c$, as the characteristic velocity and length scales, respectively, and define the dimensionless time and space coordinates $t_*=t U_\infty/c$, $x_*=x/c$, $y_*=y/c$, $\epsilon_*=\epsilon/c$, velocities $u_*=u/U_\infty$, $v_*=v/U_\infty$, pressure $p_*=p/U_\infty^2$ and vorticity $\omega_*=\omega c/U_\infty $ to obtain
\begin{equation}
    \frac{\partial \boldsymbol{u}_*}{\partial t_*}+\boldsymbol{u}_* \bcdot \nabla_* \boldsymbol{u}_*= - \nabla_* p_* 
    - \frac{C_x\boldsymbol{i} + C_y\boldsymbol{j}}{2\upi \epsilon^2_*}  e^{-(x_*^2+y_*^2)/\epsilon_*^2},    
    \label{eq:UnsteadyEuler}
\end{equation}
where $\boldsymbol{u}_*=(u_*,v_*)^\top$ is the velocity vector and $\boldsymbol{i}$ and $\boldsymbol{j}$ represent the streamwise and normal unit vectors. Here, the non-dimensional streamwise forcing, $C_x$, and normal forcing, $C_y$, model the impact of the airfoil on the flow akin to the ALM. The force coefficients are  functions of the time-dependent angle of attack $\alpha(t_*)$ and are regularised with a Gaussian kernel of width, $\epsilon_*$, centred at the actuator point located at $\boldsymbol{x}^{Act}_*=(x^{Act}_*, y^{Act}_*)^\top=(0,0)^\top$.
By taking the \textit{curl} of equation \ref{eq:UnsteadyEuler}, the pressure term can be eliminated, and one obtains a transport equation for the vorticity,
\begin{equation}
    \frac{\partial \omega_*}{\partial t_*} + \boldsymbol{u}_* \bcdot \nabla_* \omega_* = 
    \frac{-y_* C_x + x_* C_y}{\upi \epsilon^4_*}  e^{-(x_*^2+y_*^2)/\epsilon_*^2}.
    \label{eq:UnsteadyEulerVort}
\end{equation}
Equation \eqref{eq:UnsteadyEulerVort} may further be simplified by considering a linear perturbation analysis around $\omega_*= 0$, and thus $C_x=C_y=0$ such that $\boldsymbol{u}_*= \boldsymbol{i}+\boldsymbol{u}_*^p$, $\omega_*=\omega_*^p$, $C_x=C^p_x$, $C_y=C^p_y$ and $\boldsymbol{u}_* \bcdot \nabla_* \omega_* \approx \partial \omega^p_*/\partial x_*$.
After these approximations, equation \ref{eq:UnsteadyEulerVort} becomes,
\begin{equation}
      \frac{\partial \omega^p_*}
      {\partial t_*}
      + 
      \frac{\partial \omega^p_*}
      {\partial x_*}
      = \frac{-y_* C^p_x + x_* C^p_y}{\upi \epsilon^4_*}  e^{-(x_*^2+y_*^2)/\epsilon_*^2}.
      \label{eq:UnsteadyEulerVortLin}
\end{equation}
For the remainder of this paper the superscripts $^p$ are omitted for all variables, and it is understood that they refer to their respective perturbation values.
It should be noted that the linearisation point of $\omega_*=0$ implicitly assumes an infinitesimally thin non-cambered drag-free airfoil at zero mean angle of attack such that the wake of the airfoil lies on the $x$-axis at $y_*=0$ parallel to the free stream velocity $U_\infty$. Thus, applying the derived solutions to airfoils with a non-zero base loading,  camber or thickness is an approximation since the influence of mean-flow deflection on the vorticity transport is neglected. The first effect will be observed when comparing to the non-linear LES reference in Section \ref{Sec:LESvsTheory}, whereas the latter two aspects are inherently not captured by the ALM and thus also not present in the non-linear LES results. More advanced second-order approaches would be needed to take those effects into account, as for the example in the Sears problem \citep{goldstein_complete_1976, atassi_sears_1984}. The impact of the simplifying assumptions made here to arrive at the linearised vorticity transport equation \ref{eq:UnsteadyEulerVortLin} are discussed in Section \ref{Sec:NonLinError}.\par
\begin{figure}
  \centerline{\includegraphics{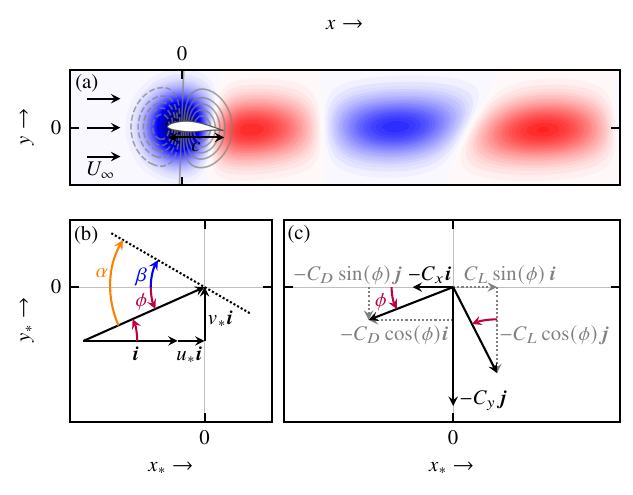}}
  \caption{(a) The problem setup with the pitching airfoil (chord $c$) represented by a two-dimensional Gaussian body force in a free stream flow of speed $U_\infty$. The resulting unsteady Gaussian forcing (right-hand-side of equation \ref{eq:UnsteadyEulerVortLin}) is illustrated by the grey dashed/solid (negative/positive forcing) contour lines. The blue/red (negative/positive vorticity) contours illustrate the resulting bound vortex and shed vorticity. Note that forcing and vorticity are time-dependent, and thus their signs and magnitude at a given spatial location can change. (b) \nolinebreak Definition of the angle of attack, $\alpha$, the pitch angle, $\beta$, the flow angle, $\phi$, and the velocity vector at the actuator point, $(x^{Act}_*, y^{Act}_*)^\top=(0,0)^\top$. (c) Definition of the streamwise $C_x$ and normal $C_y$ force coefficients and their relation to the lift $C_L$ and drag $C_D$ coefficients. Note that all force coefficients are expressed as force acting from the body on the fluid. Furthermore, the magnitude of the drag force is exaggerated to aid the visual presentation of the projection.}
\label{fig:1}
\end{figure}

Using the streamwise and normal perturbation velocities, one can define the flow angle at the actuator point $\phi=\arctan(v_*/(1+u_*))$. The flow angle is related to the angle of attack via the pitch angle $\beta$ as $\alpha=\phi+\beta$, where we consider in this work the unsteady forcing to stem from a time-dependent pitch angle, $\beta(t_*)$. The flow angle furthermore allows us to determine the streamwise and normal force coefficients from the airfoil-specific tabulated lift $C_L$ and drag $C_D$ coefficients according to the projections
\begin{align}
    C_x =& -C_L \sin(\phi) + C_D \cos(\phi)
    , \nonumber \\
    C_y =& \quad C_L \cos(\phi) + C_D \sin(\phi).
\label{eqn:projection}
\end{align} 
The problem setup together with the definitions of the coordinate system, the velocity vector, the angles and the force projection are visualised in figure \ref{fig:1}. \par
Equation \ref{eq:UnsteadyEulerVortLin}  is a linear non-homogeneous partial differential equation and can be solved using the method of characteristics (see Appendix \ref{appA}). The solution for $t_* \ge 0$ is given by
\begin{align}
    \omega_*(x_*,y_*,t_*) = 
    \,&\omega_*^{IC}(x_*-t_*,y_*)  \nonumber \\
    &+ \underbrace{\int \limits_0^{t_*} \frac{-y_* C_x(s) + (x_*+s-t_*) C_y(s)}{\upi \epsilon^4_*}  e^{-((x_*+s-t_*)^2+y_*^2)/\epsilon_*^2} \: \mathrm{d} s}_{\omega^{us}_*(x_*,y_*,t_*)}.
    \label{eq:VortSol}
\end{align}
The solution comprises two terms. The first term, $\omega_*^{IC}$, is the initial condition representing a spatio-temporal shift, $x_*-t_*=(x-U_\infty t)/c$, due to vorticity being advected by the free stream velocity, $U_\infty$, and the second term, $\omega^{us}_*$, captures the generation of shed vorticity due to the fluctuations of the unsteady right-hand-side forcing term (unsteady streamwise and normal Gaussian body forces). In principle, equation \ref{eq:VortSol} allows us to solve for the two-dimensional time-dependent vorticity field. However, if one is interested in a solution not expressed in terms of a time integral, the forcing needs to be expressed in terms of a given set of basis functions. In Section \ref{sec:AnalyticVortSol} we derive a solution by expressing the forcing in terms of a Fourier series. However, we first turn our attention towards deriving solutions for the induced velocity along the wake centre line $(x_*, y_*=0)^\top$ since the knowledge of the induced velocity at the actuator point $(x^{Act}_*, y^{Act}_*)^\top=(0,0)^\top$ is sufficient to compute the unsteady loading on the airfoil.
\section{The induced velocity due to unsteady forcing}
\label{sec:VelocitySolution}
The solution for the induced velocity at location $\boldsymbol{x}=(x_*,y_*)^\top$ due to vorticity residing at $\boldsymbol{x}^\prime=(x^\prime_*,y^\prime_*)^\top$ can be derived by applying the Biot-Savart law \citep{saffman_vortex_1992} to the general unsteady vorticity solution (equation \ref{eq:VortSol}), which for the given two-dimensional flow reads as  
\begin{equation}
    (u_*,v_*)^\top = \frac{1}{2\pi}\int \limits_{-\infty}^{+\infty}  \int \limits_{-\infty}^{+\infty} \omega_*(x_*^\prime,y_*^\prime,t_*) \frac{-(y_*-y_*^\prime)\boldsymbol{i}+(x_*-x_*^\prime)\boldsymbol{j}}{(x_*-x_*^\prime)^2+(y_*-y_*^\prime)^2} \, \mathrm{d}x_*^\prime \mathrm{d}y_*^\prime.
    \label{eq:VectorisedBiotSavart}
\end{equation}
We note that when seeking a solution restricted to the wake centre line for the linearised equations the streamwise forcing only induces streamwise velocity perturbations and the normal forcing only induces normal velocity perturbations. Hence, for each forcing direction only one velocity component of the Biot-Savart integral needs to be evaluated.

\subsection{Induced velocity due to normal forcing}
\label{sec:VelocityNormSolution}
The normal induced velocity along the wake centre line $(x_*, y_*=0)^\top$ is obtained by solving the normal component of the Biot-Savart integral given in equation \ref{eq:VectorisedBiotSavart} for a generic normal forcing $C_y(t_*)$. First, we focus here on the contribution from the second term of the unsteady vorticity solution $\omega^{us}_*$ (the time integral in equation \ref{eq:VortSol}). By switching the order of temporal and spatial integration we may obtain,
\begin{equation}
    v^{us}_*(x_*,0,t_*)=\int \limits_0^{t_*} \frac{1}{2\pi}\int \limits_{-\infty}^{+\infty}  \int \limits_{-\infty}^{+\infty} \frac{(x_*-x_*^\prime) (x_*^\prime+s-t_*)}{(x_*-x_*^\prime)^2+(-y_*^\prime)^2} \frac{C_y(s)}{\pi \epsilon_*^4} e^{-((x_*^\prime+s-t_*)^2+y_*^{\prime2})/\epsilon_*^2} \: \mathrm{d}x_*^\prime  \mathrm{d} y_*^\prime   \mathrm{d} s,
    \label{eqn:vindUnstBiotSavart}
\end{equation}
which is the normal perturbation velocity induced by the vorticity shed as a result of the unsteady forcing in the time interval $t_*\in[0,t_*]$. Integration in the two spatial directions leads to,
\begin{align}
    v^{us}_*(x_*,0,t_*)= \int \limits_0^{t_*} \frac{-C_y(s)}{2\pi} \bigg[ \frac{e^{-(x_*+s-t_*)^2/\epsilon_*^2}}{\epsilon^2} + \frac{e^{-(x_*+s-t_*)^2/\epsilon_*^2} -1}{2(x_*+s-t_*)^2} \bigg] \: \mathrm{d} s,
    \label{eqn:vindUnstSol}
\end{align}
where the complete derivation is  shown in Appendix \ref{appBnormal}. We note that the integrand in the limit $x_*+s-t_*\rightarrow 0$ tends to $-C_y(s)/(4\pi\epsilon_*^2)$, i.e., it is not singular. The derived solution shows the identical structure as the formula obtained by \cite{martinez-tossas_generalized_2024} for their generalised steady three-dimensional filtered lifting line theory.
This equivalence stems from the fact that the two problems are similar upon applying a rotation of $90^\circ$ to their coordinate system and replacing the spatial convolution integral in their solution with a spatio-temporal convolution as in equation \ref{eqn:vindUnstSol}.
So far, only the contribution from the second term of the vorticity solution $\omega^{us}_*$ has been considered. Thus the induced velocity solution in equation \ref{eqn:vindUnstSol} has to be complemented with a term $v^{IC}_*(x_*,0,t_*)$, accounting for the velocity induced by the initial vorticity field in case $\omega_*^{IC} \neq 0$ in equation \ref{eq:VortSol}. If we assume that the initial condition is given by the steady state solution of the flow over a Gaussian body force, one can take the initial condition derived by \cite{martineztossas_optimal_2017}, i.e., the velocity field induced by a Lamb-Oseen vortex with core size $\epsilon_*$ advected downstream by the background flow 
\begin{equation}
    v^{IC}_*(x_*-t_*,0) = \frac{-C_y(t_*=0)}{4\pi} \frac{1 - e^{-((x_*-t_*)/\epsilon_*)^2}}{x_*-t_*}.
    \label{eqn:vindIC}
\end{equation}
The complete solution for the normal induced velocity along the wake centre line follows by combining equations \ref{eqn:vindUnstSol} and \ref{eqn:vindIC} as
\begin{equation}
    v_*(x_*,0,t_*)=v^{IC}_*(x_*-t_*,0)+v^{us}_*(x_*,0,t_*).
    \label{eqn:completeVindSol}
\end{equation} 
While for numerical implementation purposes the structure of the solution in equation \ref{eqn:vindUnstSol} is beneficial since it does not require the calculation of the temporal gradient of $C_y(t_*)$, it is insightful to rewrite the time integral using integration by parts
\begin{equation}
    v^{us}_*(x_*,0,t_*)= -\bigg[ \frac{C_y(s)}{4\pi} \frac{1 - e^{-((x_*+s-t_*)/\epsilon_*)^2}}{x_*+s-t_*} \bigg]_{s=0}^{s=t_*} + \int \limits_0^{t_*} 
    \frac{1}{4\pi} \frac{\mathrm{d} C_y(s)}{\mathrm{d}s} \underbrace{\frac{1 - e^{-((x_*+s-t_*)/\epsilon_*)^2}}{x_*+s-t_*}}_{\varphi(x_*+s-t_*)} \: \mathrm{d} s.
    \label{eqn:vindUnstSolLambOseen}
\end{equation}
In the steady case ($\mathrm{d}C_y/\mathrm{d}s=0$), the time integral vanishes, and the term in the first brackets evaluated for the lower bound $s=0$ cancels the initial condition in equation \ref{eqn:vindIC}.  Thus, the only remaining term is the first term evaluated at the upper integration bound, which represents the induced velocity of the bound vortex, e.g., one recovers the solution of the steady problem. \par
When combining the contributions of the vorticity initial condition (equation \ref{eqn:vindIC}) and the vorticity due to unsteady forcing (equation \ref{eqn:vindUnstSolLambOseen}) to the normal induced velocity at the actuator point as $v_*(0,0,t_*)=v^{IC}_*(0-t_*,0)+v^{us}_*(0,0,t_*)$, one obtains the solution 
\begin{equation}
    v_*(0,0,t_*) = - \int_0^{t_*} \frac{1}{4\upi} \frac{\mathrm{d}C_y(s)}{\mathrm{ds}} \varphi(t_*-s) \: \mathrm{d}s,
    \label{eqn:generalVindSolution}
\end{equation}
where we exploit the fact that $\varphi(x_*+s-t_*)$ for $x_*=0$ is an anti-symmetric function and $\varphi(0)=0$. This solution is a Duhamel's integral with the indical response function $\varphi$ \citep{leishman_principles_2002}. The induced velocity at the actuator point at time $t_*$ due to the unsteady Gaussian forcing is then obtained as the superposition of all indical responses to the forcing in the time interval $s\in[0,t_*]$. \citet{wagner_uber_1925} derived an indical response function for the response of a thin airfoil to a step change of the angle of attack, which has been shown by \citet{garrick_reciprocal_1938} to be directly related to Theordorsen's function $T(k)$ in frequency domain \citep{theodorsen_general_1935}. In our case, the indical response function for an unsteady Gaussian forcing is parameterised with the Gaussian kernel width $\epsilon_*$, and from equation \ref{eqn:vindUnstSolLambOseen} it can be seen that the indical response function is given by the shape representative for the induced velocity of a Lamb-Oseen vortex with core size $\epsilon_*$ located at the streamwise location $x_*=t_*-s$.

\subsection{Induced velocity due to streamwise forcing}
\label{sec:VelocityStreamwiseSolution}
The streamwise induced velocity along the wake centre line $(x_*, y_*=0)^\top$ is obtained by solving the streamwise component of the Biot-Savart integral given in equation \ref{eq:VectorisedBiotSavart} for a generic streamwise forcing $C_x(t)$. We first focus again on the contribution from the second term of the unsteady vorticity solution $\omega^{us}_*$ (the time integral in equation \ref{eq:VortSol}). By switching the order of temporal and spatial integration, it follows that
\begin{equation}
    u^{us}_*(x_*,0,t_*)= -\int \limits_0^{t_*} \frac{1}{2\pi}\int \limits_{-\infty}^{+\infty}  \int \limits_{-\infty}^{+\infty} \frac{(-y_*^\prime)}{(x_*-x_*^\prime)^2+(-y_*^\prime)^2} \frac{-y^\prime_* C_x(s)}{\pi\epsilon_*^4} e^{-((x_*^\prime+s-t_*)^2+y_*^{\prime2})/\epsilon_*^2} \mathrm{d}x_*^\prime  \mathrm{d} y_*^\prime   \mathrm{d} s
    \label{eqn:uindUnstBiotSavart}
\end{equation}
and the two integrals in $x_*^\prime$ and $y_*^\prime$ can be solved analytically (see Appendix \ref{appBstreamwise}) to obtain 
\begin{equation}
    u^{us}_*(x_*,0,t_*)= \int \limits_0^{t_*} \frac{C_x(s)}{4\pi} \bigg[ \frac{e^{-(x_*+s-t_*)^2/\epsilon_*^2}-1}{(x_*+s-t_*)^2} \bigg] \: \mathrm{d} s.
    \label{eqn:uindUnstSol}
\end{equation}
It should be noted that in the limit case of a vanishing spatio-temporal shift $x_*+s-t_*\rightarrow0$, the integrand tends to $-C_x(s)/(4\pi\epsilon_*^2)$. \par
Furthermore, given the assumption of constant streamwise forcing $C_x \neq C_x(t_*)$, one can recover the induced velocity solution of the steady two-dimensional problem derived by \cite{martineztossas_optimal_2017} in order to show that both works are consistent. They managed to obtain the steady state solution for $x_*>>1$. We derive here the solution valid for any $x_*$, but restricted to $y_*=0$. To this end, equation \ref{eqn:uindUnstSol} is integrated for general $t_*$ but constant $C_x$, which yields
\begin{align}
    u^{us}_*(x_*,0,t_*) &= \frac{C_x}{4\pi} \bigg[ \frac{1-e^{-(x_*+s-t_*)^2/\epsilon_*^2}}{(x_*+s-t_*)} -\frac{\sqrt{\pi}}{\epsilon_*} \erf{\left(\frac{(x_*+s-t_*)}{\epsilon_*}\right)} \bigg]_{s=0}^{s=t_*} .
    \label{eqn:usteady}
\end{align}
The steady state solution along the curve $(x_*,y_*=0)^\top$ is obtained in the limit $t_*\rightarrow \infty$ using again the rule of l'H\^optial
\begin{equation}
     u^{st}_*(x_*,0) = \frac{C_x}{4\pi} \bigg[ \frac{1-e^{-x_*^2/\epsilon_*^2}}{x_*} -\frac{\sqrt{\pi}}{\epsilon_*} \erf{\left(\frac{x_*}{\epsilon_*}\right)} - \frac{\sqrt{\pi}}{\epsilon_*} \bigg].
    \label{eqn:usteady2}
\end{equation}
In the case of $x_*\rightarrow \infty$, the first term vanishes and the error function tends to one which leads to a limit value of $u^{st}_*(y_*=0)=-C_x/(2\sqrt{\pi}\epsilon_*)$ far downstream, which is exactly the result derived by \cite{martineztossas_optimal_2017}. Furthermore, at the actuator point the solution is exactly half of the limit value far downstream, i.e., $u^{st}_*(y_*=0)=-C_x/(4\sqrt{\pi}\epsilon_*)$. Similar to the normal forcing case, the solution in equation \ref{eqn:usteady2} can also be used as an initial condition in order to account for a non-zero initial vorticity field $\omega^{IC}_*(x_*-t_*,y_*)$ due to streamwise forcing, i.e.,
\begin{equation}
     u^{IC}_*(x_*-t_*,0) = \frac{C_x(t_*=0)}{4\pi} \bigg[ \frac{1-e^{-(x_*-t_*)^2/\epsilon_*^2}}{x_*-t_*} -\frac{\sqrt{\pi}}{\epsilon_*} \erf{\left(\frac{x_*-t_*}{\epsilon_*}\right)} - \frac{\sqrt{\pi}}{\epsilon_*} \bigg].
     \label{eqn:uindICSol}
\end{equation}
The complete solution for the normal induced velocity along the wake centre line then follows by combining equations \ref{eqn:uindUnstSol} and \ref{eqn:uindICSol} as
\begin{equation}
        u_*(x_*,0,t_*)=u^{IC}_*(x_*-t_*,0)+u^{us}_*(x_*,0,t_*).
        \label{eqn:completeUindSol}
\end{equation}
\section{A model for the unsteady airfoil loading}
\label{sec:FeedbackProblemSolution}
In Sections \ref{sec:VelocityNormSolution} and \ref{sec:VelocityStreamwiseSolution} 
the time-dependent solutions for the normal and streamwise induced velocity due to an unsteady Gaussian body force are derived, which allow us to define the velocity sampled at the actuator point $(x^{Act}_*, y^{Act}_*)^\top=(0,0)^\top$. These formulas will now be used to combine the velocity solutions with the forcing time histories. The forcing time histories $C_x(t_*)$ and $C_y(t_*)$ are functions of the external pitch angle input and of the induced velocities since they change the flow angle and thus, in turn, the forcing. This feedback turns the velocity solutions given by equation \ref{eqn:completeVindSol} and \ref{eqn:completeUindSol} into integral equations. The associated feedback problem can be formulated in terms of the flow angle, and its solution can be found in either the time (Section \ref{sec:TimeDomainSolution}) or frequency (Section \ref{sec:FrequencyDomainSolution}) domain in order to obtain a model for the unsteady loading of the airfoil modelled as a Gaussian body force. 

\subsection{Time domain solution: The root-finding problem}
\label{sec:TimeDomainSolution}
We now seek to obtain the time history of both the forcing and the velocity at the actuator point as a function of time using equations \ref{eqn:completeVindSol} and \ref{eqn:completeUindSol}. 
We write the time history of the velocity at the actuator point with a single equation in terms of the flow angle, as done by \citet{ning_simple_2014} and \citet{martinez_tossas_solution_2024}.
The tangent of the flow angle at the actuator point is defined by
\begin{equation}
    \tan(\phi(t_*))=\frac{v_*(0,0,t_*)}{1+u_*(0,0,t_*)},
    \label{eqn:flowAngleDefiniton}
\end{equation}
where $u_*$ and $v_*$ can both be written in terms of $\phi$; thus, this is an implicit equation where the time history, $\phi(t_*)$, is the only unknown. Rearranging equation \ref{eqn:flowAngleDefiniton} yields
\begin{equation}
    R(\phi)= v_*(0,0,t_*) \cos(\phi) - (1+u_*(0,0,t_*)) \sin(\phi)=0.
    \label{eqn:root}
\end{equation}
Given a guess for $\phi(t_*)$ as input, the following algorithm provides the steps to compute equation \ref{eqn:root}:
\begin{enumerate}
    \item Compute the angle of attack using the known pitch input, $\beta(t_*)$: $$\alpha(t_*)=\phi(t_*)+\beta(t_*).$$
    \item This effective angle of attack, which incorporates the effect of the shed vorticity, allows for the evaluation of the lift, $C_L(\alpha(t_*))$, and drag, $C_D(\alpha(t_*))$, coefficients from tabulated airfoil data, which can be subsequently converted into the corresponding streamwise and normal forcing coefficients using the projection based on the flow angle given in equation \ref{eqn:projection}:
    $$    C_x(t_*) = -C_L(t_*) \sin(\phi(t_*)) + C_D(t_*) \cos(\phi(t_*)) ,$$
    $$
    C_y(t_*) = C_L(t_*) \cos(\phi(t_*)) + C_D(t_*) \sin(\phi(t_*))
    .$$
    \item The integral equations for the induced velocities are solved via numerical integration:
        $$u_*(0,0,t_*) = u^{IC}_*(0-t_*,0) + \int \limits_0^{t_*} \frac{C_x(s)}{4\pi} \bigg[ \frac{e^{-(s-t_*)^2/\epsilon_*^2}-1}{(s-t_*)^2} \bigg] \: \mathrm{d} s,$$
        $$v_*(0,0,t_*) = v^{IC}_*(0-t_*,0) + \int \limits_0^{t_*} -\frac{C_y(s)}{2\pi} \bigg[ \frac{e^{-(s-t_*)^2/\epsilon_*^2}}{\epsilon_*^2} + \frac{e^{-(s-t_*)^2/\epsilon_*^2}-1}{2(s-t_*)^2} \bigg] \: \mathrm{d} s.$$
    \item    Finally, the residual is computed: $$R(\phi)= v_*(0,0,t_*) \cos(\phi) - (1+u_*(0,0,t_*)) \sin(\phi).$$  
\end{enumerate}
A multidimensional root-finding algorithm then iteratively evaluates steps (i)--(iv) to compute the $\phi(t_*)$ that solves $R(\phi(t_*))=0$. The algorithm above is implemented based on the python scipy.optimize.root() function \citep{virtanen_scipy_2020} using the derivative-free spectral algorithm for non-linear
equations (DF-SANE) by \citet{la_cruz_spectral_2006}.

\subsection{Frequency domain solution: The closed-loop transfer function}
\label{sec:FrequencyDomainSolution}
The time domain solution from Section \ref{sec:TimeDomainSolution} can be conveniently compared to results from LES. However, it is insightful to also formulate the feedback problem in the frequency domain since it allows for a concise analysis of the Gaussian kernel width's influence on the phase and magnitude modulation of the unsteady loading through a closed-loop transfer function. This connection is analogous to the one that can be obtained between Theodorsen's frequency domain solution for unsteady airfoil motion \citep{theodorsen_general_1935} and Wagner's time domain solution \citep{wagner_uber_1925} as shown by \cite{garrick_reciprocal_1938}. \par
\begin{figure}
  \centerline{\includegraphics{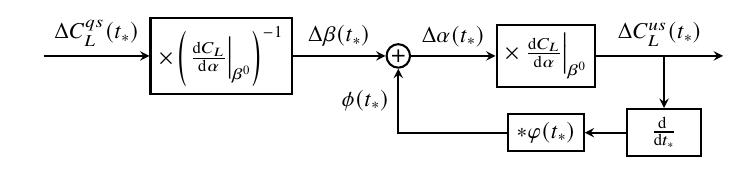}}
  \caption{Schematic block diagram of the feedback problem for the unsteady lift experienced by a pitching airfoil with $C_D=0$ at a linearised operating point $\beta^0$. The convolution operator is denoted by $\ast$, and $\varphi$ is the indical response function for the normal induced velocity (defined in equation \ref{eqn:vindUnstSolLambOseen}).}
\label{fig:blockDiagram}
\end{figure}

 
In order to derive the transfer function from the quasi-steady to the unsteady loading, we focus here on the following simplified problem of an airfoil with $C_D=0$ at an operational point $C_L(\beta^0)$ with the linear lift slope $\mathrm{d}C_L/\mathrm{d}\alpha(\beta^0)$. For a given sinusoidal pitch input signal $\beta(t_*)=\beta^0+\Delta\beta \sin(2kt_*)$ with amplitude $\Delta\beta$, the quasi-steady and unsteady lift coefficients are then, after the decay of any transient, given by
\begin{align}
    C_L^{qs}(k) &= C_L(\beta^0) + \Delta\beta \sin(2kt_*) \left.\frac{\mathrm{d}C_L}{\mathrm{d}\alpha}\right\rvert_{\beta^0},\\
    C_L^{us}(k) &= C_L(\beta^0) + \Delta\beta |G(k)| \sin(2kt_*+\angle G(k)) \left.\frac{\mathrm{d}C_L}{\mathrm{d}\alpha}\right\rvert_{\beta^0},
\end{align}
where $|G|$ and $\angle G$ are the magnitude and phase of the corresponding closed-loop transfer function $G$ mapping from the quasi-steady to the unsteady lift.  This transfer function can be derived from the system's block diagram shown in figure \ref{fig:blockDiagram}, where we explicitly invoke the small-angle approximation for the flow angle $\phi\approx v_*$. Calculating the Laplace transform of the individual blocks, the transfer function follows from block diagram algebra. It should be noted that the Laplace transform has the convenient property that the convolution of two functions in the time domain reduces simply to a multiplication in the Laplace domain. Denoting the Laplace transform of the indical response function $\varphi$ for the normal induced velocity defined in equation \ref{eqn:vindUnstSolLambOseen} as $\Phi(ik)$, the closed-loop transfer function for the feedback problem then reads as
\begin{equation}
    G(k) = \frac{\Delta C^{us}_L(k)}{\Delta C^{qs}_L(k)} = \frac{1}{1-2ki \left.\frac{\mathrm{d}C_L}{\mathrm{d}\alpha}\right\rvert_{\beta^0} \Phi(ik)},
\end{equation}
with $i$ denoting the imaginary unit (not to be confused with the streamwise unit vector $\boldsymbol{i}$). The full expression for the Laplace transform of the indical response function $\varphi$ is given by
\begin{align}
    \Phi(ik) =& \frac{1}{16\upi} \bigg[ 2 \mathbb{\gamma} - 2\upi \erfi(ik \epsilon_*) - 2\log(1/\epsilon_*^2) \nonumber \\
    &+4\log(2ki) + (2\epsilon_*ki)^2 \,_2F_2((1,1);(3/2,2);(ik\epsilon_*)^2) \bigg]
\end{align}
where $\mathbb{\gamma}$ is Euler's constant and $_pF_q(a;b;z)$ is the generalised hypergeometric function. Unsteady aerodynamics problems are commonly classified using the reduced frequency $k=\upi f c/U_\infty$, which is based on the semi-chord length \citep{leishman_principles_2002}. Recalling that the non-dimensional variables in this work are based on the chord length and the free stream velocity, it follows that $\sigma_*=2k$, which explains the added factor of 2 when writing the frequency in terms of the reduced frequency $k$. 
\par
\begin{figure}
  \centerline{\includegraphics{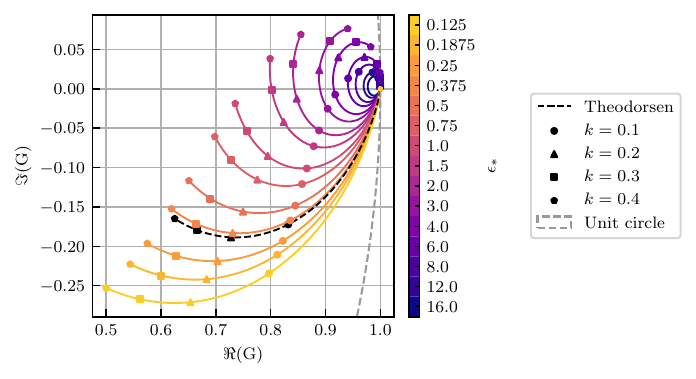}}
  \caption{The closed-loop transfer function, $G(k)$, for a pitching flat plate, ($\mathrm{d}C_L/\mathrm{d}\alpha=2\pi$), in the complex plane. The function is shown for fifteen different Gaussian kernel widths. The Theodorsen function, $T(k)$, is shown for reference \citep{theodorsen_general_1935}.}
\label{fig:transferFunctionComplexPlane}
\end{figure}

The transfer function, $G(k)$, is shown in figure \ref{fig:transferFunctionComplexPlane} in the complex plane for the case of a flat plate ($\mathrm{d}C_L/\mathrm{d}\alpha=2\upi$) up to a reduced frequency of $k=0.4$. The magnitude of the transfer function is always $|G|\leq 1$, whereas the phase can be both positive or negative depending on the considered kernel width and the reduced frequency. In general, it can be seen that smaller kernel widths tend to lead to larger damping and larger phase lags within $k\in[0,0.4]$, although these trends are not monotonous with increasing frequency. For reference, Theodorsen's function, $T(k)$, is also shown \citep{theodorsen_general_1935}. It is noted that the Theodorsen function is not the limit case of $\lim_{\epsilon \to 0} G(k)$. However, it should be stressed that there are differences in how the airfoil is represented in the two models. For the Theodorsen model, the airfoil is represented by a line distribution of singular point vortices along the chord. In contrast, the Gaussian regularisation spreads the vorticity across a surface weighted by an isotropic Gaussian. Furthermore, the present model is based on point velocity sampling. An integral velocity sampling approach as introduced by \citet{churchfield_advanced_2017} could influence the transfer function and should be explored in future work.

\section{The vorticity field created by an unsteady Gaussian body force}
\label{sec:AnalyticVortSol}
The derivations in Section \ref{sec:VelocitySolution} bypass the need to calculate an explicit vorticity solution and thus avoid the need for expressing the forcing in terms of a given set of basis functions. This approach allows the computation of the unsteady loading on an airfoil subject to a generic pitch actuation, as shown in Section \ref{sec:FeedbackProblemSolution}. However, it is instructive to study not only the influence of the Gaussian kernel width on the unsteady loading but also the unsteady vorticity field forming around the airfoil for the following two reasons: First, the vorticity field offers an explanation for the magnitude and phase modulation introduced to the unsteady loading for a given Gaussian kernel width and reduced frequency. Second, the vorticity solution can be utilised to assess the spatial and temporal nature of the modelling error caused by linearising the vorticity transport equation (equation \ref{eq:UnsteadyEulerVort}) by computing the non-linear vorticity residual with respect to the reference LES data.\par
Thus, in the following an unsteady vorticity solution for equation \ref{eq:UnsteadyEulerVortLin} is derived, which requires expressing the forcing time histories $C_x(t_*)$ and $C_y(t_*)$ as a Fourier series. Then, the solution from Section \ref{sec:TimeDomainSolution} can be utilised as a starting point to obtain the forcing time histories for a given pitch actuation signal defined on $t_*\in[0, T_*]$, incorporating the feedback effect of the shed vorticity. Subsequently, the forcing time histories are expressed as discrete Fourier series
\begin{align}
    C_x(t_*) &= C_x^0 + \sum \limits_{n=1}^N \bigg(\alpha^n_x \cos\big(2 \pi n t_*/T_*\big) + \beta^n_x \sin\big(2 \pi n t_*/T_*\big) \bigg), \\
    C_y(t_*) &= C_y^0 + \sum \limits_{n=1}^N \bigg(\alpha^n_y \cos\big(2 \pi n t_*/T_*\big) + \beta^n_y \sin\big(2 \pi n t_*/T_*\big) \bigg),
\end{align}
where $\alpha^n_x$, $\beta^n_x$, $\alpha^n_y$ and $\beta^n_y$ are the real Fourier coefficients for the streamwise and normal forcing, respectively. By exploiting the linearity of equation \ref{eq:UnsteadyEulerVortLin}, one may consider the solution for each Fourier basis function separately and obtain the complete solution by superposition. To this end, consider the combined streamwise and normal forcing term (right-hand side of equation \ref{eq:UnsteadyEulerVortLin}) for a single frequency component to be
\begin{align}
    F^n(x_*,y_*, t_*) = &\frac{-y_* (\alpha^n_x \cos(\sigma_*^n t_*) + \beta^n_x \sin(\sigma_*^n t_*))}{\pi \epsilon_*^4} e^{-(x_*^2+y_*^2)/\epsilon_*^2} \nonumber \\
    &+ \frac{x_* (\alpha^n_y \cos(\sigma_*^n t_*) + \beta^n_y \sin(\sigma_*^n t_*))}{\pi \epsilon_*^4} e^{-(x_*^2+y_*^2)/\epsilon_*^2},
    \label{eqn:forcingFourierSeries}
\end{align}
where $\sigma_*^n=2\pi n t_*/T_*$ is the harmonic angular frequency.\par
The complete vorticity solution is then obtained by superposing three contributions: the unsteady vorticity contributions from each non-zero frequency $\omega^{n}_*$, the contributions $\omega^{0}_*$ from the zero frequency components $C_x^0$ and $C_y^0$, and, if applicable, the contribution from the initial vorticity field $\omega^{IC}_*$, which in this work is considered to be the vorticity field corresponding to the steady state solution (equations (8) and (15) from \cite{martineztossas_optimal_2017}).  In the following, we derive the missing solutions for the first two contributions for normal and streamwise forcing and assemble the complete solutions.

\subsection{Vorticity due to normal forcing}
The vorticity solution accounting for the unsteady non-zero frequency contributions ($n>0$) of the normal forcing is obtained by inserting the normal forcing component of equation \ref{eqn:forcingFourierSeries} into the integral in equation \ref{eq:VortSol} to obtain
\begin{equation}
    \omega^{n}_*=\int \limits_{x_*-t_*}^{x_*} \xi \frac{\alpha^n_y \cos[\sigma_*^n (\xi-x_*+t_*)] + \beta^n_y \sin[\sigma_*^n (\xi-x_*+t_*)]}{\pi \epsilon_*^4} e^{-(\xi^2+y_*^2)/\epsilon_*^2} \mathrm{d} \xi,
    \label{eqn:CyForcingIntegral}
\end{equation}
where the transformation $\xi=x_*+(s-t_*)$ ($\mathrm{d}\xi=\mathrm{d}s$) is used. The analytical solution for this integral is derived in Appendix \ref{appCnorm}. The solution corresponding to the zero frequency component of the unsteady forcing ($n=0$) is obtained by inserting the forcing $C^0_y$ instead of the trigonometric functions into the integral in equation \ref{eq:VortSol}, which then can readily be solved to obtain
\begin{equation}
    \omega^{0}_* = -\frac{C^0_y}{2\pi\epsilon_*^2}e^{-(x_*^2+y_*^2)/\epsilon_*^2} + \frac{C^0_y}{2\pi\epsilon_*^2} e^{-[(x_*-t_*)^2+y_*^2]/\epsilon_*^2}.
\end{equation}
Finally, all three vorticity contributions can be combined to end up with the vorticity solution given an unsteady normal forcing $C_y(t_*)$ and subject to the assumption that $C_y(t_*)$ can be expressed as a discrete Fourier series:
\begin{align}
     \omega(x_*,y_*,t_*) =& \omega^{IC}_* + \omega^{0}_* + \omega^{n}_* \nonumber \\
    =& -\frac{C^{IC}_y}{2\pi\epsilon_*^2} e^{-[(x_*-t_*)^2+y_*^2]/\epsilon_*^2} \nonumber \\
    &-\frac{C^0_y}{2\pi\epsilon_*^2}e^{-(x_*^2+y_*^2)/\epsilon_*^2} + \frac{C^0_y}{2\pi\epsilon_*^2} e^{-[(x_*-t_*)^2+y_*^2]/\epsilon_*^2} \nonumber \\
     &+ \sum \limits_{n=1}^N \left\{ - \frac{\alpha^n_y \cos(\sigma_*^n t_*) + \beta^n_y \sin(\sigma_*^n t_*)}{2\pi\epsilon_*^2}e^{-(x_*^2+y_*^2)/\epsilon_*^2} \right. \nonumber \\
     & \quad \quad \quad \left. + \frac{\alpha^n_y}{2\pi\epsilon_*^2} e^{-[(x_*-t_*)^2+y_*^2]/\epsilon_*^2} \right. \nonumber\\
     & \quad \quad \quad \left. + \frac{\sigma_*^n}{4\pi\epsilon_*^2} e^{-y_*^2/\epsilon_*^2} \left\{ (\alpha^n_y i+\beta^n_y) e^{i\sigma_*^n(-x_*+t_*)} \left(-\frac{1}{2}i\sqrt{\pi}\epsilon_* e^{-1/4 (\sigma_*^n)^2 \epsilon_*^2}\right) \right. \right. \times \nonumber\\
     & \quad \quad \quad \left. \left. \quad \bigg[ \erfi\left(\frac{\sigma_*^n\epsilon_*}{2}+\frac{ix_*}{\epsilon_*}\right) - \erfi\left(\frac{\sigma_*^n\epsilon_*}{2}+\frac{i(x_*-t_*)}{\epsilon_*}\right) \bigg] \right. \right. \nonumber\\
     & \quad \quad \quad +\left. \left.  (-\alpha^n_y i+\beta^n_y) e^{-i\sigma_*^n(-x_*+t_*)} \left(\frac{1}{2}i\sqrt{\pi}\epsilon_* e^{-1/4 (\sigma_*^n)^2 \epsilon_*^2}\right) \right. \right. \times \nonumber\\
     & \quad \quad \quad \left. \left. \quad \bigg[ \erfi\left(\frac{\sigma_*^n\epsilon_*}{2}-\frac{ix_*}{\epsilon_*}\right) - \erfi\left(\frac{\sigma_*^n\epsilon_*}{2}-\frac{i(x_*-t_*)}{\epsilon_*}\right) \bigg] \right\} \right\}.
    \label{eq:AnalyticVortSolutionCy}
\end{align}
It can be noted that in the case of a constant forcing and an already established bound vortex (i.e., $C^{IC}_y=C_y^0$) the solution simply reduces to the steady state solution derived in \cite{martineztossas_optimal_2017}. 

\subsection{Vorticity due to streamwise forcing}
The vorticity solution accounting for the unsteady non-zero frequency contributions ($n>0$) of the streamwise forcing is obtained by inserting the streamwise forcing component of equation \ref{eqn:forcingFourierSeries} into the integral in equation \ref{eq:VortSol}. Making use again of the transformation $\xi=x_*+(s-t_*)$, the integral reads as
\begin{equation}
    \omega^n_* = -\frac{y_*}{\pi \epsilon_*^4} \int \limits_{x_*-t_*}^{x_*} (\alpha^n_x \cos[\sigma_*^n (\xi-x_*+t_*)] + \beta^n_x \sin[\sigma_*^n (\xi-x_*+t_*)])e^{-(\xi^2+y_*^2)/\epsilon_*^2} \mathrm{d} \xi,
    \label{eqn:CxForcingIntegral}
\end{equation}
and its solution is derived in Appendix \ref{appCstream}. The solution corresponding to the zero frequency component ($n=0$) of the unsteady forcing is obtained by inserting the forcing $C^0_x$ instead of the trigonometric functions into the integral in equation \ref{eq:VortSol}, which can then be solved to obtain
\begin{equation}
    \omega^{0}_* = \frac{C^0_x}{2\sqrt{\upi} \epsilon_*^3} y_* e^{-y_*^2/\epsilon_*^2} \bigg[ \erf\left(\frac{x_*-t_*}{\epsilon_*}\right) - \erf\left(\frac{x_*}{\epsilon_*}\right)\bigg].
\end{equation}
Combining all three contributions leads to the complete vorticity solution resulting from periodic streamwise forcing $C_x(t_*)$:
\begin{align}
     \omega(x_*,y_*,t_*) = & \omega^{IC}_* + \omega^{0}_* + \omega^{n}_* \nonumber \\ 
                         = & - \frac{C^{IC}_x}{2\sqrt{\upi} \epsilon_*^3} y_* e^{-y_*^2/\epsilon_*^2} \bigg[ 1 + \erf\left(\frac{x_*-t_*}{\epsilon_*}\right)\bigg] \nonumber \\
                         & + \frac{C^0_x}{2\sqrt{\upi} \epsilon_*^3} y_* e^{-y_*^2/\epsilon_*^2} \bigg[ \erf\left(\frac{x_*-t_*}{\epsilon_*}\right) - \erf\left(\frac{x_*}{\epsilon_*}\right)\bigg]  \nonumber \\
                         &+ \sum \limits_{n=1}^N \left\{-\frac{y_*}{2\pi\epsilon_*^4} e^{-y_*^2/\epsilon_*^2} \left\{ (\alpha^n_x -i\beta^n_x) e^{i\sigma_*^n(-x_*+t_*)} \left(-\frac{1}{2}i\sqrt{\pi}\epsilon_* e^{-1/4 (\sigma_*^n)^2 \epsilon_*^2}\right) \right. \right. \times \nonumber\\
                         &  \quad \quad \quad \left. \left. \quad \bigg[ \erfi\left(\frac{\sigma_*^n\epsilon_*}{2}+\frac{ix_*}{\epsilon_*}\right) - \erfi\left(\frac{\sigma_*^n\epsilon_*}{2}+\frac{i(x_*-t_*)}{\epsilon_*}\right) \bigg] \right. \right.  \nonumber\\
                         &  \quad \quad \quad +\left. \left.  (\alpha^n_x+i\beta^n_x) e^{-i\sigma_*^n(-x_*+t_*)} \left(\frac{1}{2}i\sqrt{\pi}\epsilon_* e^{-1/4 (\sigma_*^n)^2 \epsilon_*^2}\right) \right. \right. \times \nonumber\\
                         &  \quad \quad \quad \left. \left. \quad \bigg[ \erfi\left(\frac{\sigma_*^n\epsilon_*}{2}-\frac{ix_*}{\epsilon_*}\right) - \erfi\left(\frac{\sigma_*^n\epsilon_*}{2}-\frac{i(x_*-t_*)}{\epsilon_*}\right) \bigg] \right\} \right\}.
    \label{eq:AnalyticVortSolutionCx}
\end{align}
An observation similar to the normal forcing case can be made. In the case of constant forcing and an already established vorticity (i.e., $C^{IC}_x=C_x^0$), the solution simply reduces to the steady state solution derived in \cite{martineztossas_optimal_2017}.

\FloatBarrier
\section{Validation: Model versus numerical simulation}
\label{Sec:LESvsTheory}
In the previous four sections we have derived a model for the unsteady loading on an airfoil modelled as a two-dimensional Gaussian body force. In this section, we validate this model by comparing with LES data to establish the bounds of its accuracy and model limitations.

\subsection{Investigated cases}
\begin{figure}
  \centerline{\includegraphics{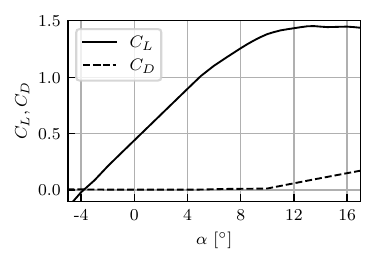}}
  \caption{Tabulated lift, $C_L(\alpha)$, and drag, $C_D(\alpha)$, coefficients of the NACA64-A17 airfoil.}
\label{fig:TabulatedAirfoilData}
\end{figure}

The range of investigated cases is chosen such that a range of operating points, types of actuation, reduced frequencies and the interplay between streamwise and normal forcing are explored. All cases are conducted using the NACA64-A17 airfoil, which is, for example, also used for the outer part of the blade of the NREL 5MW reference turbine \citep{jonkman_definition_2009}. The airfoil's tabulated lift and drag coefficients are shown in figure \ref{fig:TabulatedAirfoilData}.
Each investigated case is analysed for the set of Gaussian kernel widths $\epsilon_* \in \{0.25, 0.5, 1.0, 2.0, 4.0\}$. The lower limit is chosen here on the order of the optimal kernel width \citep{martineztossas_optimal_2017}, while the upper limit corresponds to coarse grid ALM-LES simulations conducted for wind energy purposes. An overview of all cases and their acronyms is provided in table \ref{tab:cases}. In the following, the different cases are explained in more detail. 
\begin{table}
    \centering
    \begin{tabular}{cccccccc}
        Case acronym & Operating & \multicolumn{2}{c}{Forcing} & Actuation type & Magnitude $\Delta \alpha$ & Amplitude $\Delta \beta$ & Frequency $k$\\
         & point $\beta^0$ & $C_x$ & $C_y$ & & (Step) & (Periodic) & (Periodic)\\
        A0-Cx-S4 & $0^\circ$ & Yes & No & Step & $4^\circ$ & - & -\\
        A0-Cxy-S4 & $0^\circ$ & Yes & Yes & Step & $4^\circ$ & - & -\\
        A8-Cx-S12 & $8^\circ$& Yes & No & Step & $8^\circ$ & - & -\\
        A8-Cxy-S12 & $8^\circ$& Yes & Yes & Step & $8^\circ$ & - & -\\
        A14-Cxy-S18 & $14^\circ$& Yes & Yes & Step & $18^\circ$ & - & -\\
        A0-Cxy-P3-k01 & $0^\circ$ & Yes & Yes & Step + Periodic & $4^\circ$ & $3^\circ$ & $k=0.1$\\
        A0-Cxy-P3-k02 & $0^\circ$ & Yes & Yes & Step + Periodic & $4^\circ$ & $3^\circ$ & $k=0.2$\\
        A0-Cxy-P3-k03 & $0^\circ$ & Yes & Yes & Step + Periodic & $4^\circ$ & $3^\circ$ & $k=0.3$\\
    \end{tabular}
    \caption{Overview of the investigated cases. The initial condition for all cases is $\omega^{IC}=0$,  corresponding to a situation without initial forcing. The step magnitude is defined as the difference between the pitch angle after the step (given by the operating point $\beta^0$) and the angle of attack, which results in approximately zero lift force for the cambered airfoil $\Delta \alpha = \beta^0-\alpha^{C_L=0}$ ($\alpha^{C_L=0}\approx-4^\circ$). The periodic component of the actuation is of the form $\beta(t_*)=\Delta \beta \sin(2kt_*)$. The step components contain a continuous frequency spectrum where the magnitude of the Fourier transform of the step scales as $|\hat{\beta}| \propto 1/k$.}
    \label{tab:cases}
\end{table}
\par
In order to characterise the obtained theoretical solutions and their scaling with the kernel width, we begin by considering two cases with a step in pitch angle. Both cases start from an initial condition of zero vorticity, $\omega^{IC}=0$, i.e., zero forcing. This choice ensures that there is no mismatch between the initial conditions used for the model and LES. The simulation start thus corresponds to a sudden step actuation to reach the operating pitch angle $\beta^0$ for $t_*\geq 0$.  The considered operating points are $\beta^0=0^\circ$ (A0-Cxy-S4) and $\beta^0=8^\circ$ (A8-Cxy-S12). The same cases are also repeated without applying any normal forcing (A0-Cx-S4 and A8-Cx-S12). It should be noted that the frequency content of a step function is proportional to the inverse of the frequency ($|\hat{\beta}|\propto 1/k$) and thus tests the model simultaneously across the entire frequency range. The step case A14-Cxy-S18 is only conducted for $\epsilon_*=0.25$ to obtain the largest forcing magnitudes. This case is used to perform a grid and domain size convergence study for the LES setup.\par
The second set of cases considers the same initial step but additionally superposes continuous periodic pitch actuation with an amplitude of $\Delta \beta=3^\circ$ and reduced frequencies of $k=0.1$, $k=0.2$ and $k=0.3$  (indicated by the endings -k01, -k02 and -k03 of the case acronyms). The periodic pitch actuation is of the form $\beta(t_*)=\Delta \beta \sin(2kt_*)$. It is applied at the operating point $\beta^0=0^\circ$ (A0-Cxy-P3-k...). This second set of cases is well suited to test the predictions of the derived closed-loop transfer function for an isolated frequency component after the initial transient decayed. 

\subsection{Employed numerical code and simulation setup}
\label{subsec:LESsetup}
Numerical non-linear reference results for the assessment of the developed model are obtained by means of  LES using the open-source code AMR-Wind (available at \url{https://github.com/Exawind/amr-wind}). AMR-Wind solves the three-dimensional incompressible Navier-Stokes equations on block-structured Cartesian grids and is specifically designed for wind energy applications. For details on the numerical schemes employed in AMR-Wind, the reader is referred to  \citet{almgren_conservative_1998} and \citet{sharma_exawind_2024}.  For the purpose of validating the developed model, which is based on the Euler equations, the governing equations are solved in the infinite Reynolds number limit by setting the molecular viscosity to zero. The inflow and the wake flow of interest are laminar, and thus no subgrid-scale model is employed. \par
This work utilises AMR-Wind's implementation of the ALM for straight translating wings. AMR-Wind does not allow for strictly two-dimensional simulations; rather, we simulate the airfoil in a three-dimensional domain as infinite wing, i.e., periodic boundary conditions are imposed in the spanwise direction. The lower and upper boundary condition in the normal direction  is a slip wall ($y$-direction). In the streamwise direction ($x$-direction), a constant inflow velocity of $\boldsymbol{u}=U_{\infty}\boldsymbol{i}$ is imposed at the upstream  boundary (which in non-dimensional units corresponds to $\boldsymbol{u}_*=\boldsymbol{i}$ ), and a pressure outflow boundary condition is applied at the downstream boundary. The wing spans two actuator points located on the spanwise boundaries of the domain at which the actuator forces are applied. The forces are regularised with a two-dimensional isotropic Gaussian kernel of width $\epsilon_*$. The velocity used for the actuator force calculation is the known inflow velocity (see Appendix \ref{appD}).  \par
\begin{figure}
  \centerline{\includegraphics[width=13cm]{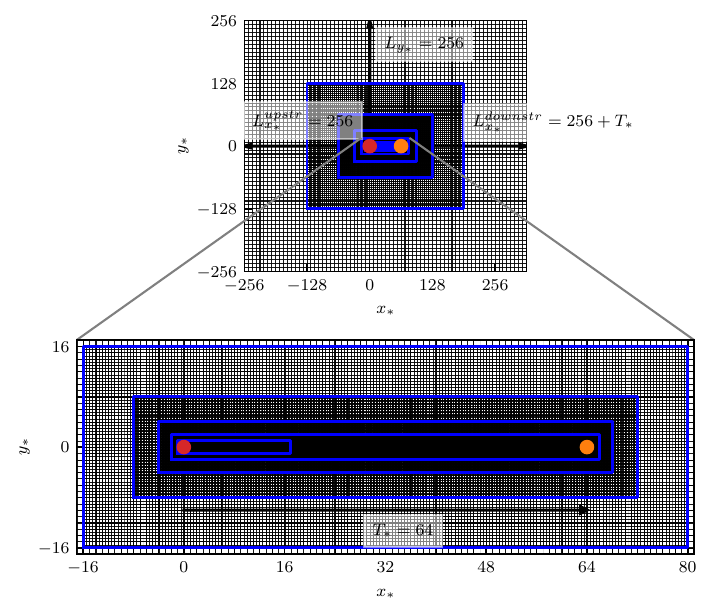}}
  \caption{LES domain and grid refinement structure for the Gaussian kernel width $\epsilon_*=0.25$ with the resolution on the finest grid level being $\epsilon_*/\Delta x=8$. Note that the domain here is configured for a simulation time of $T_*=64$ in order to keep the aspect ratio limited for visual clarity. The actual simulations are performed for $T_*=128$ (steps) and  $T_*=256$ (periodic). The red dot marks the actuator point. The orange dot marks the approximate downstream location of the start-up vortex at time $t_*=T_*$.}
\label{fig:LesDomainSetup}
\end{figure}

The convergence of LES results with domain size and grid resolution are studied in Appendix \ref{appE} for the A14-Cxy-S18 case using $\epsilon_*=0.25$. Based on this convergence study, all simulations are conducted with a semi-domain height $L_{y_*}=256$ and an upstream fetch of $L^{upstr}_{x_*}=256$ between the inflow boundary and the actuator point. The streamwise distance from the actuator point to the outlet is $L^{downstr}_{x_*}=T_*+L_{y_*}$, where $T_*$ is the maximum simulated time. The grid resolution on the coarsest grid level is constant for all simulations; however, depending on the studied kernel width, four ($\epsilon_*=4$) to eight  ($\epsilon_*=0.25$) refinement levels are added such that on the finest grid around the actuator point the resolution is always $\epsilon_*/\Delta x_*=8$.  The grid resolution between two adjacent grid levels varies with a constant factor of 2.  The size of the finest grid level is dependent on the kernel width. Its extent in the $x$ and $y$-directions is $-4\epsilon_*< x_* < 16+4\epsilon_*$ and $-4\epsilon_*< y_* < 4\epsilon_*$, respectively. Letting the index $m$ denote the grid level (the finest being $m=0$), the extent of all other refinement levels ($m>0$) in $x$ and $y$ is given by $-4\epsilon_* 2^m< x_* < T_*+4\epsilon_* 2^m$ and $-4\epsilon_* 2^m< y_* < 4\epsilon_* 2^m$.
The time step size $\Delta t_{*,\text{LES}}$ is chosen such that the Courant-Friedrichs-Levy (CFL) number defined on the finest grid level is $CFL<0.5$ at all times. Step cases are simulated up to $T_*=128$, whereas the step cases followed by periodic actuation are simulated up to $T_*=256$. These long simulation times provide enough time for the actuator loading to reach the new steady state (step cases) or periodic limit cycle (step+periodic cases) after the decay of the initial transient. The simulation time-dependent streamwise domain length ensures that no vorticity is truncated at the outlet. The simulation time for the periodic cases is chosen twice as long in order to collect a large number of actuation periods, even for the smallest frequency $k=0.1$. The LES setup is visualised in figure \ref{fig:LesDomainSetup}.

\subsection{Response to pitch step actuation}
\label{sec:resultsStepResponse}
We start by probing the developed model for two pitch step actuation cases within the linear regime of the lift curve. Their respective operating points are $\beta^0=0^\circ$ (A0-Cxy-S4) and $\beta^0=8^\circ$ (A8-Cxy-S12). The initial condition is chosen as undisturbed free stream flow, i.e., no forcing is active. Consequently, the two operating points correspond to effective angle of attack steps of $\Delta \alpha=4^\circ$ and $\Delta \alpha=12^\circ$ upon simulation start (the lift coefficient of the cambered airfoil is approximately zero at $\alpha\approx -4^\circ$).\par

\begin{figure}
  \centerline{\includegraphics{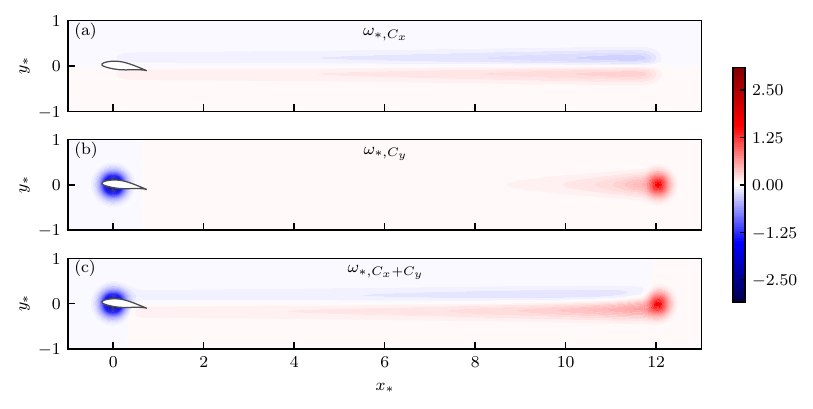}}
  \caption{Model solution for the vorticity field of the A8-Cxy-S12 case with the kernel width $\epsilon_*=0.25$ at time instance $t_*=12$. The vorticity created by streamwise $C_x$ and normal $C_y$ forcing is shown in (a) and (b), respectively. The total resultant vorticity field is shown in (c). The airfoil modelled by the Gaussian body force is shown in white with black outline. The airfoil's quarter-chord point is located at the actuator point $(x^{Act}_*, y^{Act}_*)^\top=(0,0)^\top$.}
\label{fig:analyticalVortField_tSnap12_eps0.25_A8-Cxy-S12}
\end{figure}

The vorticity distribution around the airfoil and in the wake as predicted by the model, with $\epsilon_*=0.25$ is shown in figure \ref{fig:analyticalVortField_tSnap12_eps0.25_A8-Cxy-S12} for the larger pitch step case A8-Cxy-S12. Around the airfoil, a bound vortex forms in response to the non-zero loading after $t_*>0$. The bound vortex is accompanied by the formation of a start-up vortex, which ensures conservation of the overall vorticity. Since the model assumes the advection speed in the wake to match the free stream velocity, the start-up vortex is located at $x_*=12$ in the wake at the shown time instance $t_*=12$. Furthermore, it is seen that the wake centre line $(x_*, y_*=0)^\top$ is the symmetry axis for both the vorticity due to streamwise and the vorticity due to normal forcing (figure \ref{fig:analyticalVortField_tSnap12_eps0.25_A8-Cxy-S12}a/b). The vorticity due to streamwise forcing exhibits an anti-symmetric distribution around this symmetry axis, whereas the normal forcing contribution is distributed symmetrically around the wake centre line. These distributions stem from the symmetry properties of the forcing terms in the linearised vorticity transport equation \ref{eq:UnsteadyEulerVortLin}.

\begin{figure}
  \centerline{\includegraphics{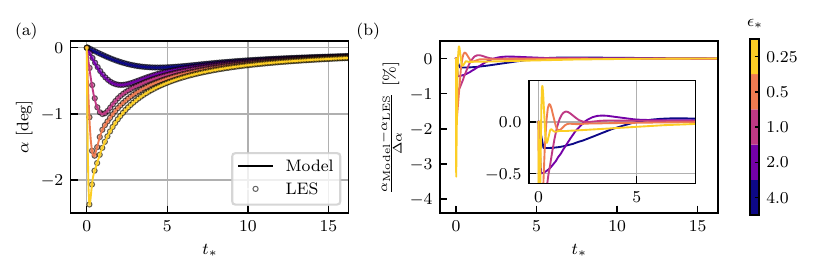}}
  \caption{Comparison of the model solution and LES reference for the initial time evolution of the angle of attack (a) and the corresponding relative error (b) for the pitch step actuation case A0-Cxy-S4. Reference data points are only shown in steps of $4\Delta t_{*,\text{LES}}/\epsilon_*$.}
\label{fig:aoa_A0-Cxy-S4}
\end{figure}

The formation of the start-up vortex and its subsequent advection downstream causes a dynamic angle of attack response. A comparison of the model and the LES solution for this response is shown in figure \ref{fig:aoa_A0-Cxy-S4} for the small pitch step case A0-Cxy-S4. The start-up vortex induces velocity at the actuator point, which in turn alters the angle of attack. As it is advected further downstream, its influence on the angle of attack fades, which leads the angle of attack to converge towards the new steady state operating point $\alpha=\beta^0=0^\circ$ (figure \ref{fig:aoa_A0-Cxy-S4}a). The initial modulation of the angle of attack depends on the kernel width since it sets the local strength of the applied forcing and thus also the strength of the start-up vortex. Furthermore, the core size of the start-up vortex is also set by the kernel width. The core size determines the time period between $t_*=0$ and the time instance when the angle of attack reaches its minimum. The minimum angle of attack is attained faster with smaller kernel width and thus core size.\par
\begin{figure}
  \centerline{\includegraphics{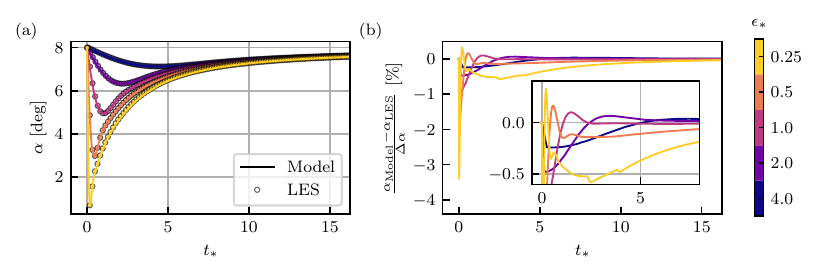}}
  \caption{Comparison of the model solution and LES reference for the initial time evolution of the angle of attack (a) and the corresponding relative error (b) for the pitch step actuation case A8-Cxy-S12. Reference data points are only shown in steps of $4\Delta t_{*,\text{LES}}/\epsilon_*$.}
\label{fig:aoa_A8-Cxy-S12}
\end{figure}

The model error compared to the LES reference is presented in figure \ref{fig:aoa_A0-Cxy-S4}b. The very initial relative angle of attack error is as large as $3.4\%$ and then subsequently drops to the order of tenths of a percent. The relative error starts approaching zero after the actuator point no longer lies within the vortex core region of the start-up vortex. The relative error is larger for smaller kernel widths as long as there is overlap between the vortex core and the actuator point. 
The larger pitch step case A8-Cxy-S12 shown in figure \ref{fig:aoa_A8-Cxy-S12} follows the previous observations made apart from the relative error evolution for $\epsilon_*=0.25$. In this case, the error also persists after the vortex core no longer overlaps with the actuator point and only fades at $t_*\approx 10$. The larger pitch step combined with the smallest kernel width makes for a stronger bound vortex, increasingly challenging the model, which is based on a linearisation around the vorticity free background flow. 


\begin{figure}
  \centerline{\includegraphics{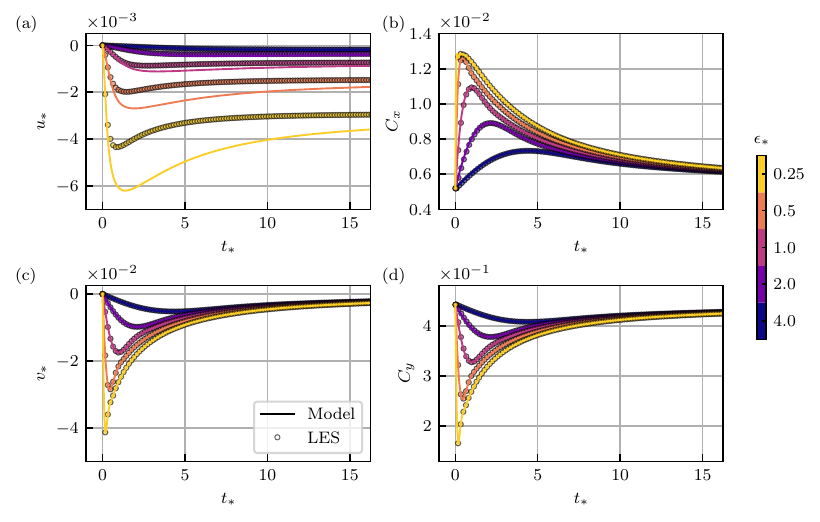}}
  \caption{Comparison of the model solution and LES reference for the initial time evolution of the streamwise (a) and normal (c) induced velocity and the streamwise (b) and normal (d) force coefficients for the pitch step actuation case A0-Cxy-S4.  Reference data points are only shown in steps of $4\Delta t_{*,\text{LES}}/\epsilon_*$.}
\label{fig:indVel_A0-Cxy-S4}
\end{figure}

\begin{figure}
  \centerline{\includegraphics{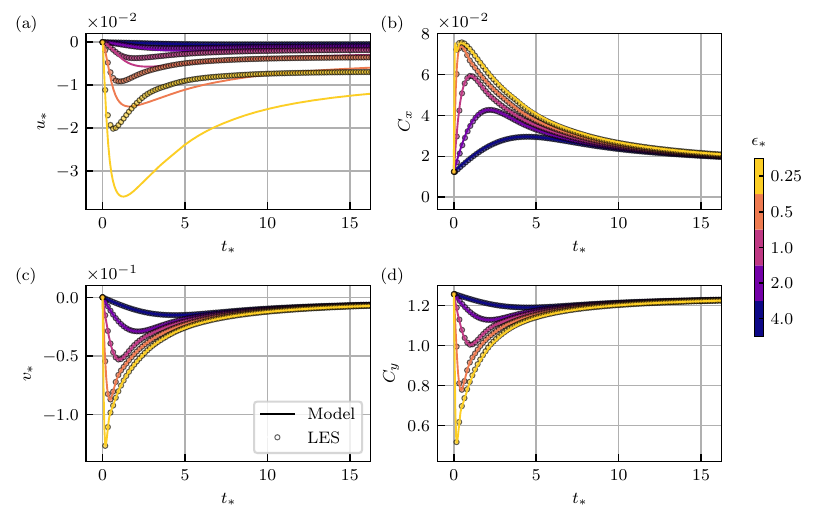}}
  \caption{Comparison of the model solution and LES reference for the initial time evolution of the streamwise (a) and normal (c) induced velocity and the streamwise (b) and normal (d) force coefficients for the pitch step actuation case A8-Cxy-S12.  Reference data points are only shown in steps of $4\Delta t_{*,\text{LES}}/\epsilon_*$.}
\label{fig:indVel_A8-Cxy-S12}
\end{figure}

We explore the two pitch step cases further by turning our attention towards the loading, which the model ultimately should predict. The model is able to accurately reproduce the time evolution of the streamwise, $C_x$, and normal, $C_y$, force coefficients for both the small (figure \ref{fig:indVel_A0-Cxy-S4}b/d) and the large (figure \ref{fig:indVel_A8-Cxy-S12}b/d) pitch step cases. In particular, the transient drop in normal forcing and peak in streamwise forcing caused by the start-up vortex are captured.
It is also worthwhile to study the induced velocities at the actuator point, which induce the change in flow angle and thus angle of attack (panels a/c of figures \ref{fig:indVel_A0-Cxy-S4} and \ref{fig:indVel_A8-Cxy-S12}). 
The model matches the LES prediction for the normal induced velocity, $v_*$, for both pitch step magnitudes, but discrepancies are apparent for the streamwise induced velocity, $u_*$, which become more pronounced for smaller kernel widths and the larger pitch step (compare figures \ref{fig:indVel_A0-Cxy-S4}a and \ref{fig:indVel_A8-Cxy-S12}a). This discrepancy also matches the previous observation that the angle of attack error of the A8-Cxy-S12 case for $\epsilon_*=0.25$ persists for a longer time. 
Since we deal with small perturbation velocities, the flow angle is small and thus approximately given by $\phi\approx v_*$. Thus, the error in streamwise induced velocity does not affect the quality of the streamwise and normal force coefficients as long as the small perturbation assumption is justified.
In fact, $u_*$ is a measure of the system's non-linearity and is relevant for the flow angle calculation given larger induced velocities. However, larger induced velocities start to challenge the linearisation of the vorticity transport equation for the model derivation. Thus, the mismatch for $u_*$ and its increase for stronger forcing is consistent. In Section \ref{Sec:NonLinError} we show that the observed streamwise velocity residual is explained by the vorticity residual obtained from the difference of the model (linear) and the LES reference (non-linear) solution. However, the two pitch step cases show that the developed model can predict the loading on the airfoil in the linear regime of the lift curve despite the mismatch in $u_*$. The cases also illustrate the importance of the kernel width. The largest studied kernel width of $\epsilon_*=4.0$ barely modifies the normal force coefficient while the kernel width $\epsilon_*=0.25$, which is of the order of the optimal value, results in a transient drop of more than $50\%$ (see figure \ref{fig:indVel_A8-Cxy-S12}d).


\begin{figure}
  \centerline{\includegraphics{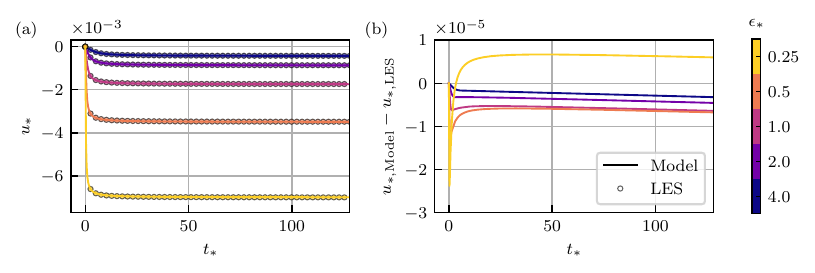}}
  \caption{Comparison of the model solution and LES reference for the time evolution of the streamwise induced velocity (a) and the corresponding absolute error (b) for the pitch step actuation case A8-Cx-S12. Reference data points are only shown in steps of $64\Delta t_{*,\text{LES}}/\epsilon_*$.}
\label{fig:uInd_A8-Cx-S12}
\end{figure}

Motivated by the mismatch of the streamwise induced velocity we also conduct the pitch step cases with an absent normal force $C_y=0$ (the cases labelled A0-Cx-S4 and A8-Cx-S12). The resulting evolution of $u_*$ and the corresponding absolute model error are shown in figure \ref{fig:uInd_A8-Cx-S12} for the large pitch step case. In the absence of normal forcing, the previously observed mismatch between model and the LES reference vanishes and the model error after the initial time steps is 2 orders of magnitude smaller than $u_*$. 
This highlights the impact of the assumptions made when linearising the vorticity transport equation for the model derivation. Neglecting the non-linear transport terms, which include the induced velocity, causes the model to miss the normal displacement of the start-up vortex as it is advected in the induced normal velocity field of the bound vortex, which is maximal at the wake centre line $(x_*,y_*=0)$. This lack of vertical displacement is also apparent in the previously shown model solution for the vorticity field (figure \ref{fig:analyticalVortField_tSnap12_eps0.25_A8-Cxy-S12}). Switching off the normal forcing, $C_y$, removes its anti-symmetric contribution to the induced velocity field. Thus, as long as the streamwise induced velocity remains small, the model can now correctly predict $u_*$, too. It can also be noted that the induced streamwise velocity after the decay of the transient tends towards the steady state limit given by $u_*=-C_x/(4\sqrt{\pi}\epsilon_*)$, which was also derived by \cite{martineztossas_optimal_2017}.

\subsection{Response to periodic pitch actuation}
\label{sec:resultsPeriodicResponse}
After characterising the developed model with pitch steps we now move on towards signals which are more present in reality, i.e., smooth signals and, in particular, periodic signals with sinusoidal shape. The solution to the latter can also be used to construct the solution for more complex forcing signals via their representation in terms of a Fourier series. In the following, we consider the three cases A0-Cxy-P3-k01, A0-Cxy-P3-k02 and A0-Cxy-P3-k03, which initially possess the same pitch step as the A0-Cxy-S4 case but then proceed with a continuous sinusoidal pitch actuation of amplitude $\Delta \beta=3^\circ$. The three different cases vary in terms of the actuation frequency; namely, the three reduced frequencies $k=0.1$, $k=0.2$ and $k=0.3$ are considered. \par

\begin{figure}
  \centerline{\includegraphics{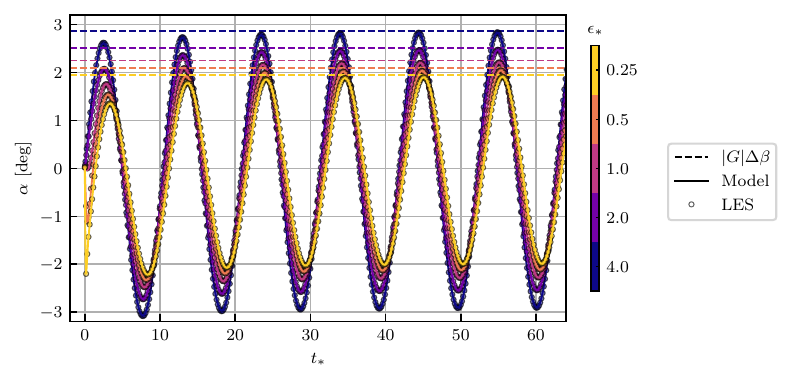}}
  \caption{Comparison of the model solution and LES reference for the initial time evolution of the angle of attack for the periodic pitch actuation case A0-Cxy-P3-k03. Reference data points are only shown in steps of $4\Delta t_{*,\text{LES}}/\epsilon_*$. The grey dotted line marks the amplitude of the pitch actuation ($\Delta \beta=3^\circ$). The horizontal coloured dashed lines indicate the unsteady angle of attack amplitudes $|G(k)|\Delta \beta$ predicted by the model closed-loop transfer function $G(k)$.}
\label{fig:aoa_A0-Cxy-P3-k03}
\end{figure}

Representative angle of attack evolutions obtained for these three reduced frequencies are shown for the case of $k=0.3$ in figure \ref{fig:aoa_A0-Cxy-P3-k03}. The angle of attack now varies periodically where its initial evolution shows a transient due to the influence of the start-up vortex, which initially is still present. After the transient, the angle of attack attains a limit cycle. For reference, figure \ref{fig:aoa_A0-Cxy-P3-k03} shows horizontal lines indicating the angle of attack amplitude predicted by the model closed-loop transfer function (given by $|G(k)|\Delta \beta$). The frequency domain solution in terms of the transfer function is able to predict the limit cycle amplitude modulation since the peaks of the angle of attack time series converge towards the horizontal lines. Therefore, the effect of increasingly smaller kernel widths is a successive reduction of the angle of attack amplitude compared to the quasi-steady reference ($\Delta \beta=3^\circ$) and a successive increase in phase lag. For the given airfoil and reduced frequency of $k=0.3$, the reduction in effective angle of attack amplitude is as large as $35\%$ for the smallest studied kernel width $\epsilon_*=0.25$, which is of the order of the optimal kernel width. 

\begin{figure}
  \centerline{\includegraphics{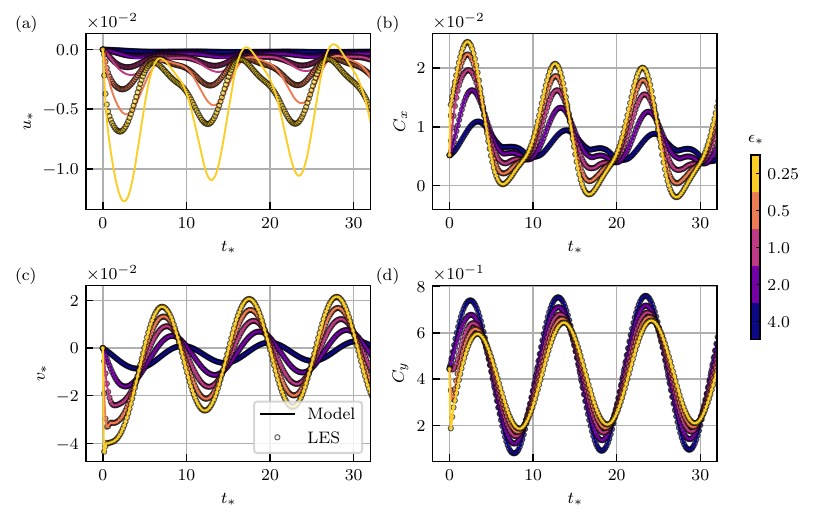}}
  \caption{Comparison of the model solution and LES reference for the initial time evolution of the streamwise (a) and normal (c) induced velocity and the streamwise (b) and normal (d) force coefficients for the periodic pitch actuation case A0-Cxy-P3-k03.  Reference data points are only shown in steps of $4\Delta t_{*,\text{LES}}/\epsilon_*$.}
\label{fig:indVel_A0-Cxy-P3-k03}
\end{figure}

The evolution of the induced velocities and the force coefficients corresponding to the previous angle of attack evolution of the A0-Cxy-P3-k03 case are shown in figure \ref{fig:indVel_A0-Cxy-P3-k03}. Similar to the step cases, the model cannot predict the correct streamwise induced velocity, $u_*$, in the presence of combined streamwise and normal forcing, where the error increases with decreasing kernel width.

\begin{figure}
  \centerline{\includegraphics{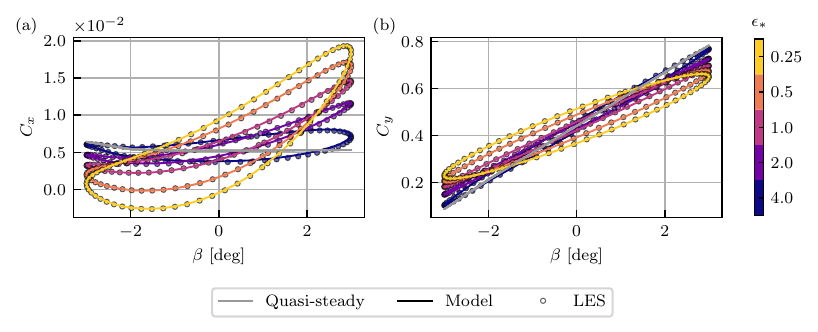}}
  \caption{Comparison of the model solution and LES reference for the hysteresis of the streamwise (a) and normal (b) force coefficient (periodic pitch actuation case A0-Cxy-P3-k03). The quasi-steady references $C_x(\beta)$ and $C_y(\beta)$ are shown in grey.}
\label{fig:hysteresis_A0-Cxy-P3-k03}
\end{figure}

After the decay of the initial transient, it is convenient to represent the force coefficients in terms of hysteresis plots (figure \ref{fig:hysteresis_A0-Cxy-P3-k03}). The grey reference lines in this plot show the quasi-steady force coefficients, i.e., the forcing as it would be obtained when the instantaneous angle of attack would match the instantaneous pitch angle. However, the velocity induced by the shed vorticity in the wake causes a non-zero flow angle, and it is $C_y(\beta)\neq C_y(\alpha)$. The degree to which these two coefficients deviate depends on the kernel width. For the shown case of $k=0.3$ in figure \ref{fig:hysteresis_A0-Cxy-P3-k03}b, it can be seen that smaller kernel widths cause stronger hysteresis, e.g., the minor axis of the ellipse for $C_y$ is larger (phase modulation). Furthermore, the major axis of the ellipse becomes increasingly tilted towards the horizontal axis, i.e., the force amplitude is damped compared to the quasi-steady case (amplitude modulation). For the streamwise forcing, $C_x$, smaller kernel widths also lead to more pronounced hysteresis as observed for $C_y$ (figure \ref{fig:hysteresis_A0-Cxy-P3-k03}a). However, the trend for the amplitude modulation is different compared to $C_y$ in the sense that the unsteady solution for $C_x$ compared to the quasi-steady reference shows an increase in amplitude rather than a decrease. This amplitude increase is larger for smaller kernel widths because for these cases, the normal induced velocity reaches larger values (figure \ref{fig:indVel_A0-Cxy-P3-k03}c), and thus the same follows for the flow angle $\phi$. Since the streamwise force coefficient in the linear regime of the lift curve is dominated by the $C_L$ contribution, it attains its maximum when the product $-C_L(\alpha) \sin(\phi)$ becomes maximal, which in addition to large $C_L$ also requires flow angles close to their maximal negative value since $C_L$ and $v_*$ are approximately half a period out of phase (compare panels c and d of figure \ref{fig:indVel_A0-Cxy-P3-k03}).

\begin{figure}
  \centerline{\includegraphics{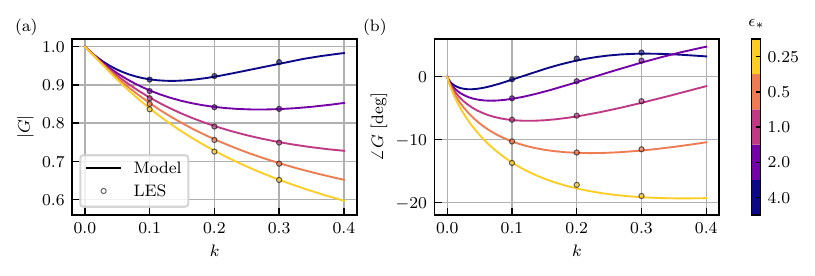}}
  \caption{The magnitude (a) and phase (b) of the model closed-loop transfer function $G(k)$ for the NACA64-A17 airfoil at the linearisation point $\beta^0=0^\circ$. Dots indicate the corresponding amplitude and phase modulations extracted from the last period of the LES reference cases A0-Cxy-P3-k01, A0-Cxy-P3-k02 and A0-Cxy-P3-k03.}
\label{fig:frequDomain_lesVersusModel}
\end{figure}

Given the dominance of the normal forcing in the linear regime of the lift curve for which the present model is developed, the observations regarding the phase and amplitude modulations made from the time series and hysteresis curves can be concisely summarised in a Bode plot. Figure \ref{fig:frequDomain_lesVersusModel}a/b show the Bode magnitude and Bode phase plot of the model closed-loop transfer function $G(k)$, which was derived assuming $C_D=0$. $G(k)$ is here computed using the linearised value of the lift slope of the NACA64-A17 airfoil at the operating point $\beta^0=0^\circ$, matching the operating point of the periodic actuation cases. The model solution is presented for the set of kernel widths $\epsilon_* \in \{0.25, 0.5, 1.0, 2.0, 4.0\}$ in the range $k\in[0,0.4]$. In addition, the phase and amplitude modulation is extracted from the last period of the LES reference data for the cases A0-Cxy-P3-k01 ($k=0.1$), A0-Cxy-P3-k02 ($k=0.2$) and A0-Cxy-P3-k03 ($k=0.3$), which are shown as dots in the plot. The transfer function solution matches the LES reference for all three reduced frequencies. \par

For the studied reduced frequencies $k\leq0.3$, magnitude and phase trends are monotone for varying kernel width at a given constant frequency. However, the magnitude and phase variation with frequency for a constant kernel width are not monotone. The magnitude reaches a minimum, after which it again approaches $|G|=1$ for further increasing $k$. The convergence to $|G|=1$ happens for smaller $k$, the larger the kernel width, and for the shown relevant frequency range, it is only visible for $\epsilon_*=4.0$. In this limit the unsteady lift magnitude again approaches the quasi-steady solution. Also, the phase tends towards a common limit of $\angle G=0^\circ$ for all kernel widths as $k\rightarrow \infty$, although this convergence happens for $k>0.4$ for the studied set of kernel widths. Consequently, for a chosen kernel width representing the airfoil, the maximum reduction of the quasi-steady lift amplitude $\text{min}(|G(k; \epsilon_*)|)$ and maximum phase lag $\text{min}(\angle G(k; \epsilon_*))$ occur for an intermediate reduced frequency $0<k<\infty$. Further, the maximum phase lag occurs for smaller $k$ than does the maximum amplitude reduction. Within the range of relevant $k$, smaller kernel widths lead to larger damping and increased phase lag. Comparing $\epsilon_*=0.25$, which is of the order of the optimal kernel width, with $\epsilon_*=4$, which corresponds to a value encountered for coarse grid ALM-LES for wind energy purposes, one would miss about $\Delta(|G|)\approx0.3$ of damping and $\Delta (\angle G) \approx 23^\circ$ of phase lag at a reduced frequency of $k=0.3$. This difference is the unsteady equivalent of the steady state lift error observed for suboptimal large kernel widths, which motivated the development of smearing corrections \citep{meyer_forsting_vortex-based_2019, martinez-tossas_filtered_2019,  dag_new_2020, kleine_non-iterative_2023}. \par

\begin{figure}
  \centerline{\includegraphics{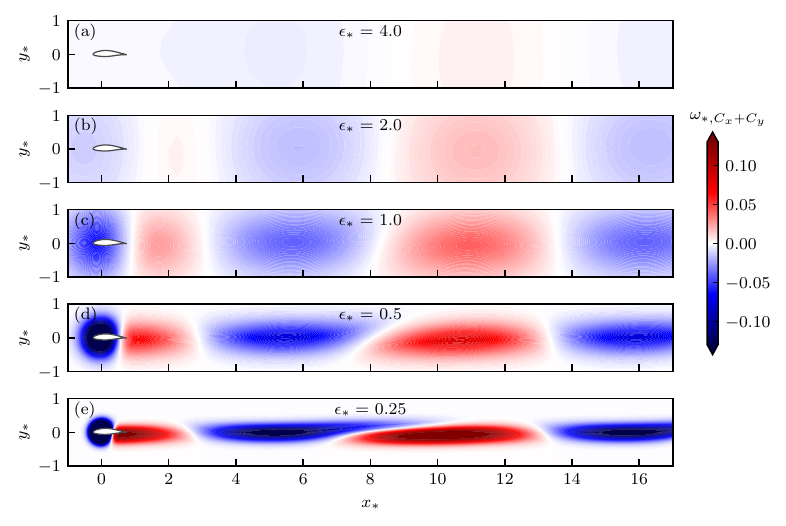}}
  \caption{Model solution for the total resultant vorticity field of the A0-Cxy-P3-k03 case at time instance $t_*=32$. The panels (a) to (e) show the solution for the Gaussian kernel widths $\epsilon \in \{4.0, 2.0, 1.0, 0.5, 0.25\}$. The airfoil modelled by the Gaussian body force is shown in white with black outline. The airfoil's quarter-chord point is located at the actuator point $(x^{Act}_*, y^{Act}_*)^\top=(0,0)^\top$. }
\label{fig:compareAnalyticalVorticityFields_tSnap32_A0-Cxy-P3-k03}
\end{figure}
The source of the unsteady error due to large kernel widths can also be made apparent by comparing the topology of the shed vorticity in the wake for varying $\epsilon_*$ (figure \ref{fig:compareAnalyticalVorticityFields_tSnap32_A0-Cxy-P3-k03}). The regularisation with the Gaussian kernel conserves both the integrated force as well as the integrated vorticity. That is assuming the magnitude of the point force would not be affected by the induced velocities, which in reality it is. But even neglecting this effect, the conservation on an integral level can still be achieved by locally very different vorticity distributions. The vorticity for small kernel widths is more localised, and thus integral conservation directly implies locally higher vorticity magnitudes in proximity of the airfoil and the wake centre line. In turn, large kernel widths spread the vorticity across a larger area, reducing its local magnitude. Since the induced velocity scales with the inverse of the distance outside of the vortex core, this directly implies smaller induced velocities at the actuator point. Finally, the phase error is also apparent from the vorticity distributions. To this end, one can track the vertical band of zero vorticity, which is located between the bound vortex (blue) and the neighbouring positive vorticity patch (red) directly downstream of the airfoil. Its location is closest to the airfoil for $\epsilon_*=0.25$ and then shifts continuously farther downstream for larger kernel widths, i.e., the larger the kernel width, the further ahead the phase.\par
Summarising the model validation presented in this section, we conclude that the developed model is capable of predicting the unsteady loading on the airfoil within the considered operating regime, which is the linear part of the lift curve. The step and periodic actuation cases demonstrated the frequency-dependent impact of the Gaussian kernel width on the unsteady loading both in terms of lift reduction and added phase lag.
\section{Discussion: Assessing the model error}
\label{Sec:NonLinError}
The preceding validation results show that within the linear regime of the lift curve, the derived model solutions are capable of predicting the unsteady loading of the airfoil. However, it also became clear that even within this regime the model cannot reproduce the correct streamwise induced velocity in the presence of combined forcing in the streamwise and normal directions. This deficiency does not affect the accuracy of the results as long as the small angle approximation for the flow angle is valid. Nevertheless, due to this observation we already pointed in Section \ref{sec:resultsStepResponse} towards the fact that there is a coupling between the two forcing directions, which is neglected for the derivations in this work. We derive here the exact form of the neglected terms and assess their magnitude and spatial localisation with respect to the reference LES data.\par
The vorticity transport equation for the two-dimensional problem was introduced in equation \ref{eq:UnsteadyEulerVort} and reads in expanded form as
\begin{equation}
    \frac{\partial \omega_*}{\partial t_*} + \frac{\partial \omega_*}{\partial x_*} + \underbrace{u_* \frac{\partial \omega_*}{\partial x_*} + v_* \frac{\partial \omega_*}{\partial y_*} }_{\mathcal{N}(\boldsymbol{u}_*, \omega_*)} = 
    \underbrace{\frac{-y_* C_x + x_* C_y}{\upi \epsilon^4_*}  e^{-(x_*^2+y_*^2)/\epsilon_*^2}}_{F(x_*,y_*,t_*)}.
    \label{eq:VortTransAllTermsDecomposed}
\end{equation}
In order to arrive at the final linearised equation \ref{eq:UnsteadyEulerVortLin} as introduced in Section \ref{Sec:UnsteadyGaussianBodyForces}, one needs to assume that the non-linear term $\mathcal{N}(\boldsymbol{u}_*, \omega_*)$ can be neglected, which means that if the other remaining terms are all of order $\mathcal{O}(1)$, the non-linear term should be at maximum of order $\mathcal{N}=\mathcal{O}(1/10)$. An increasing magnitude of the forcing $F(x_*,y_*,t_*)$ increasingly challenges this linearisation. Analysis of the forcing shows that it is locally linearly proportional to the magnitude of the force coefficients $F \propto C_i$ ($i\in{x,y}$), but inversely proportional to the fourth power of the kernel width $F \propto 1/\epsilon_*^4$. However, in addition the Gaussian contribution, the forcing term attains a maximum value of $\text{max}(x_*\exp(-x_*^2/\epsilon_*^2))\approx 0.43 \epsilon_*$ at $x=\epsilon_*/\sqrt{2}$. Neglecting the effect of the kernel on the strength of the forcing via the induced velocity, this scaling shows that within the linear region of the lift curve, doubling the angle of attack only doubles the local forcing magnitude, whereas halving the Gaussian kernel width leads to an eightfold increase in local forcing (i.e. $F\propto 1/\epsilon_*^3$).\par

\begin{figure}
  \centerline{\includegraphics{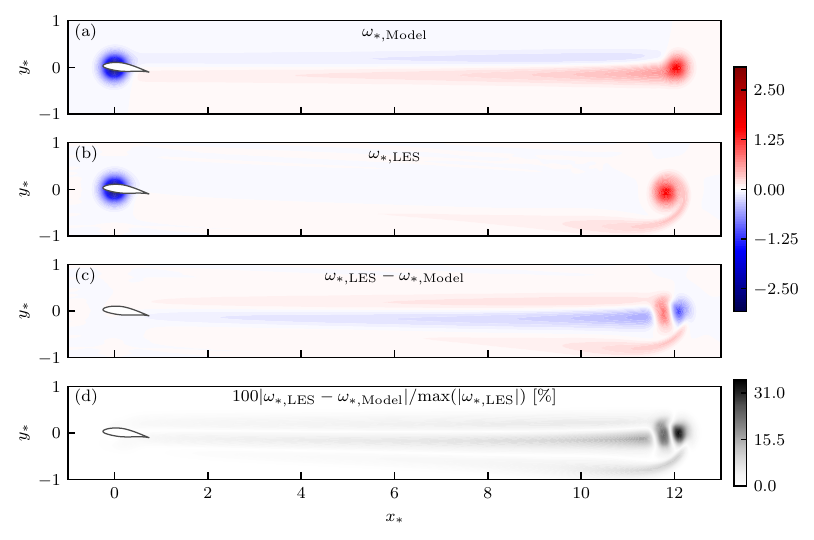}}
  \caption{Model (a) and LES (b) solution for the total vorticity field of the A8-Cxy-S12 case for $\epsilon_*=0.25$ at time instance $t_*=12$. The difference between the model and LES solution is shown in (c). The relative error is shown in (d) where the reference for the normalisation of the error is the maximum value of the absolute vorticity in the $x_*-y_*$ plane. The airfoil modelled by the Gaussian body force is shown with a grey outline. The airfoil's quarter-chord point is located at the actuator point $(x^{Act}_*, y^{Act}_*)^\top=(0,0)^\top$. }
\label{fig:lesVStheoryVortField_tSnap12_eps0.25_A8-Cxy-S12}
\end{figure}

\begin{figure}
  \centerline{\includegraphics{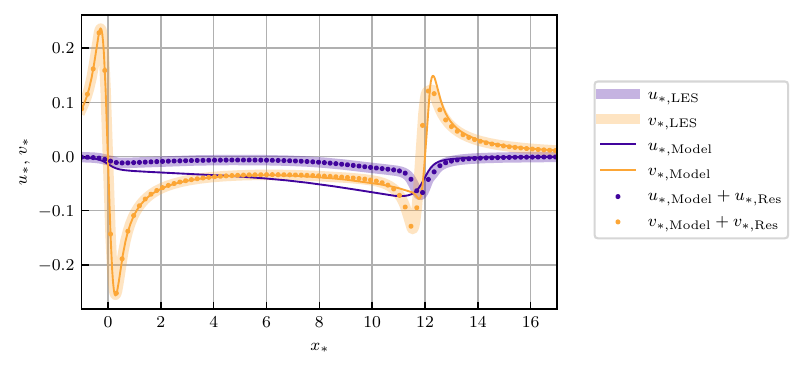}}
  \caption{Comparison of the model and the LES solution for the streamwise and normal induced velocity along the curve $(x_*, y_*=0)^\top$ for $\epsilon_*=0.25$ at time instance $t_*=12$ (A8-Cxy-S12 case). The black markers show the sum of the model solution and the velocity induced by the vorticity residual shown in figure\ref{fig:lesVStheoryVortField_tSnap12_eps0.25_A8-Cxy-S12}c. The residual contribution is obtained by numerical evaluation of the Biot-Savart integral.}
\label{fig:lesVStheoryVelField_tSnap12_eps0.25_A8-Cxy-S12}
\end{figure}

Based on this scaling analysis. we focus the error analysis on the smallest kernel width of $\epsilon_*=0.25$. Figure \ref{fig:lesVStheoryVortField_tSnap12_eps0.25_A8-Cxy-S12}a/b shows the vorticity solution for the large pitch step case A8-Cxy-S12 obtained both from the model and the LES at time instance $t_*=12$. The LES which solves for all terms present in equation \ref{eq:VortTransAllTermsDecomposed} predicts a roll-up and normal displacement of the start-up vortex. Furthermore, the non-linear terms break the (anti-) symmetric vorticity distribution of the model with respect to the wake centre line. In contrast, the model cannot capture these effects since it neglects $\mathcal{N}$. Figure \ref{fig:lesVStheoryVortField_tSnap12_eps0.25_A8-Cxy-S12}c/d shows the resulting (relative) error with respect to the LES reference. In proximity of the start-up vortex, the maximum relative error reaches more than $30\%$, but it remains of the order of a few percent in the remaining wake region between bound and start-up vortex. We now show that this vorticity error accounts for the discrepancy of the streamwise induced velocity at the actuator point, which is observed in figure \ref{fig:indVel_A8-Cxy-S12}a. To this end, we numerically evaluate the Biot-Savart law of the vorticity residual $\omega_{*,\text{LES}}-\omega_{*,\text{Model}}$ to obtain the velocity residuals $u_{*,\text{Res}}$ and $v_{*,\text{Res}}$. Figure \ref{fig:lesVStheoryVelField_tSnap12_eps0.25_A8-Cxy-S12} shows the streamwise and normal induced velocity along the wake centre line both for the model and the LES. The model correctly predicts $v_*$ near the actuator point and in the near wake, but deviations occur farther downstream near the start-up vortex. The streamwise component $u_*$ also deviates already in the proximity of the actuator point as observed previously. However, adding the numerically computed residual velocities to the model solution recovers the LES solution. This finding shows that the derived model solutions for the induced velocity (equations \ref{eqn:vindUnstSol} and \ref{eqn:uindUnstSol}) and the vorticity (equations \ref{eq:AnalyticVortSolutionCy} and \ref{eq:AnalyticVortSolutionCx}) are consistent and that neglecting the term $\mathcal{N}$ explains the observed model error.


\begin{figure}
  \centerline{\includegraphics{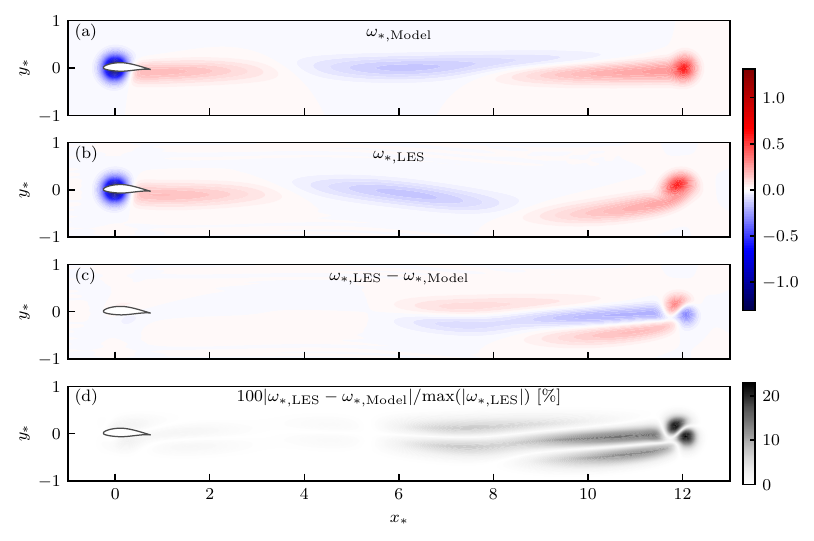}}
  \caption{Model (a) and LES (b) solution for the total vorticity field of the A0-Cxy-P3-k03 case for $\epsilon_*=0.25$ at time instance $t_*=12$. The difference between the model and LES solution is shown in (c). The relative error is shown in (d) where the reference for the normalisation of the error is the maximum value of the absolute vorticity in the $x_*-y_*$ plane. The airfoil modelled by the Gaussian body force is shown with a grey outline. The airfoil's quarter-chord point is located at the actuator point $(x^{Act}_*, y^{Act}_*)^\top=(0,0)^\top$. }
\label{fig:lesVStheoryVortField_tSnap12_eps0.25_A0-Cxy-P3-k03}
\end{figure}
\begin{figure}
  \centerline{\includegraphics{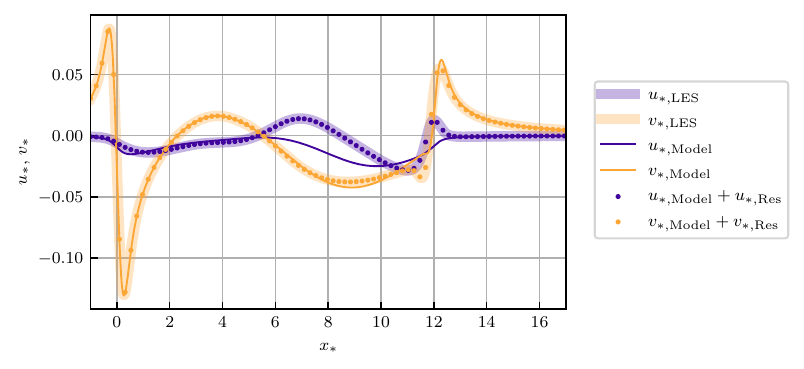}}
  \caption{Comparison of the model and the LES solution for the streamwise and normal induced velocity along the curve $(x_*, y_*=0)^\top$ for $\epsilon_*=0.25$ at time instance $t_*=12$ (A0-Cxy-P3-k03 case). The black markers show the sum of the model solution and the velocity induced by the vorticity residual shown in figure\ref{fig:lesVStheoryVortField_tSnap12_eps0.25_A0-Cxy-P3-k03}c. The residual contribution is obtained by numerical evaluation of the Biot-Savart integral.}
\label{fig:lesVStheoryVelField_tSnap12_eps0.25_A0-Cxy-P3-k03}
\end{figure}
The same error analysis is also conducted for the A0-Cxy-P3-k03 case in figure \ref{fig:lesVStheoryVortField_tSnap12_eps0.25_A0-Cxy-P3-k03}, which features a continuous change of bound vorticity and thus also a continuous band of shed vorticity. The (relative) error shown in panels c and d of this figure is constrained to a region starting downstream of $x_*=6$. This localisation shows that the previously observed absence of large errors in the near wake of the pitch step case are not simply due to the absence of vorticity (the start-up vortex was already advected further downstream), but that the linearised vorticity transport equation is capable of predicting an accurate trajectory of the shed vorticity in the near wake. This observation also manifests in the induced velocity distributions along the centre line shown in figure \ref{fig:lesVStheoryVelField_tSnap12_eps0.25_A0-Cxy-P3-k03}. The model captures the normal induced velocity accurately up to $x_*=6$. The deviations farther downstream and in general for the streamwise induced velocity are again shown to be accounted for by the contribution of the vorticity residual between LES and the model. 

\FloatBarrier
\section{Conclusions}
\label{Sec:Conclusions}
In this study, we assessed the ability of the actuator line model to capture the unsteady aerodynamic effects of induced velocity due to shed vorticity.
We focused on the simplified two-dimensional problem of an airfoil represented by an unsteady Gaussian body force. The objective was to determine the unsteady loading on the airfoil in response to generic unsteady pitch actuation with an explicit dependence on the Gaussian kernel width. This problem formulation thereby links the ALM to the theories developed by  \citet{theodorsen_general_1935} and \citet{wagner_uber_1925} for the unsteady inviscid thin-airfoil problem. A linearised vorticity transport equation was obtained from the Euler equations subject to an unsteady Gaussian body force, which enabled the derivation of theoretical solutions for the induced velocity along the wake centre line and the complete two-dimensional vorticity field. The former solution takes the form of a Duhamel's integral involving an indical response function. Based on the theoretical velocity solutions, a model for the unsteady airfoil loading was derived. Its solution was obtained in the time domain by solving a root-finding problem for the flow angle as well as in the frequency domain by deriving the system's closed-loop transfer function mapping from the quasi-steady to the unsteady lift force. The closed-loop transfer function is parameterised by the Gaussian kernel width and thus explicitly shows its impact on the amplitude modulation and phase lag of the unsteady lift as a function of the reduced frequency.\par
The model was validated with non-linear reference data obtained by means of LES. The validation considered both pitch step and periodic pitch actuation cases, where the model was tested within the linear region of the airfoil's lift curve. The relative errors for the angle of attack were on the order of tenths of a percent. Within the considered range of kernel widths $\epsilon_*\in[0.25, 4.0]$ and validated reduced frequencies $k\in\{0.1,0.2,0.3\}$, the difference between the smallest and largest kernel width on the amplitude damping can be as large as $\Delta|G|\approx0.3$, whereas the effect on the phase lag is small ($\Delta \angle G \approx 23^\circ$). The model was shown to accurately capture the time evolution of the induced normal velocity, the angle of attack and the streamwise and normal force coefficients within the linear regime of the lift curve. However, a mismatch between LES and the model was observed for the streamwise induced velocity in the case of combined streamwise and normal forcing. This mismatch was shown to be a direct consequence of the linearisation of the vorticity transport equation during the model derivation by computing the induced velocity corresponding to the vorticity residual between the non-linear LES results and the theoretical vorticity solution. \par
Based on the results, we have identified three key directions for future work. Firstly, the model derivation is currently based on a linearisation point equivalent to undisturbed (vorticity free) flow. Incorporating a flexible linearisation point by using the corresponding steady-state solution could improve the model capabilities of predicting the trajectory of the shed vorticity in the wake. A second path is to extend the model to three dimensions, e.g., to obtain an unsteady three-dimensional filtered lifting line solution, extending the steady-state solution derived by \citet{martinez-tossas_filtered_2019}. Finally, this unsteady three-dimensional filtered lifting line solution could then be employed as a correction for coarse grid ALM-LES.

\backsection[Acknowledgements]{E.T. would like to thank the Wind Energy Science group at the National Renewable Energy Laboratory (NREL) for enabling his research stay at NREL in Boulder, during which most progress on this work was made and for their warm hospitality. E.T.'s fellow PhD colleagues David Fidalgo Domingos, Marcus Becker and Daniel van den Berg are thanked for their ideas which ultimately lead to the development of the frequency domain solution of the model.}

\backsection[Funding]{This work is part of the Hollandse Kust Noord wind farm innovation program where CrossWind C.V., Shell, Eneco and Siemens Gamesa are teaming up; funding for the PhDs and PostDocs was provided by CrossWind C.V. and Siemens Gamesa. We further would like to acknowledge SURF for the computational resources made available on the Dutch
national supercomputer Snellius (grant number: EINF-6784). \\
This work was authored in part by the National Renewable Energy Laboratory, operated by Alliance for Sustainable Energy, LLC, for the U.S. Department of Energy (DOE) under Contract No. DE-AC36-08GO28308. Funding provided by the U.S. Department of Energy Office of Energy Efficiency and Renewable Energy Wind Energy Technologies Office. The views expressed in the article do not necessarily represent the views of the DOE or the U.S. Government. The U.S. Government retains and the publisher, by accepting the article for publication, acknowledges that the U.S. Government retains a nonexclusive, paid-up, irrevocable, worldwide license to publish or reproduce the published form of this work, or allow others to do so, for U.S. Government purposes.
}

\backsection[Declaration of interests]{The authors report no conflict of interest.}

\backsection[Author ORCIDs]{E. Taschner, https://orcid.org/0000-0002-0296-8168; G. Deskos, https://orcid.org/0000-0001-7592-7191; M. Kuhn, https://orcid.org/0000-0002-5506-7777; J.-W. van Wingerden, https://orcid.org/\linebreak0000-0003-3061-7442; L.A. Mart{\'i}nez-Tossas, https://orcid.org/0000-0003-2353-4999}

\backsection[Author contributions]{E.T, G.D., J.-W.W. and L.M. derived the model solutions. E.T. and L.M. implemented the model solutions and post-processing code. M.K. and E.T. made the necessary changes to AMR-Wind to enable the LES reference simulations. E.T. set up and performed the LES simulations with the help of L.M. and M.K. All authors contributed to analysing data, reaching conclusions and writing the paper.}

\appendix

\section{Solving the linearised vorticity transport equation}\label{appA}
In order to solve equation \ref{eq:UnsteadyEulerVortLin}, one can exploit its advection character by considering solutions along spatio-temporal curves given by $X(s)=x_*+s$ and $T(s)=t_*+s$ (neglecting the $y_*$ dependence, which is not relevant for the derivation). Differentiation of the solution along these curves yields
\begin{equation}
    \frac{d}{ds}\omega_*(X(s),T(s))=\frac{\partial \omega_*}{\partial x}\frac{dX}{ds}+\frac{\partial \omega_*}{\partial t}\frac{dT}{ds}=\frac{\partial \omega_*}{\partial x}+\frac{\partial \omega_*}{\partial t}=F(X(s),T(s))
\end{equation}
where $F(X(s),T(s))$ represents the right-hand side of equation \ref{eq:UnsteadyEulerVortLin}.  Subsequent integration from $T(s=-t_*)=0$ to $T(s=0)=t_*$ provides
\begin{equation}
    \int \limits_{-t_*}^0 \frac{d}{ds}\omega_*(X(s),T(s)) \: \mathrm{d}s = \omega_*(x_*,t_*)-\omega_*(x_*-t_*,0)=\int \limits_{-t_*}^0 F(x_*+s,t_*+s) \: \mathrm{d}s.
\end{equation}
Using the transformation $s^\prime=t_*+s$ ($ds^\prime=ds$) and subsequently denoting $s^\prime$ again as $s$, the solution with the explicit $y_*$ dependence is obtained as shown in equation \ref{eq:VortSol}.

\section{Derivation of the induced velocity solutions}\label{appB}
\subsection{Normal induced velocity}\label{appBnormal}
Starting from equation \ref{eqn:vindUnstBiotSavart}, integration of the portion of the integral with $y_*^{\prime}$ dependence yields 
\begin{equation}
    \int \limits_{-\infty}^{+\infty} \frac{(x_*-x_*^\prime)}{(x_*-x_*^\prime)^2+(-y_*^\prime)^2} e^{-y_*^{\prime2}/\epsilon_*^2}  \: \mathrm{d} y_*^\prime = \text{sgn}(x_*-x_*^\prime) \pi \bigg[1-\erf\left(\frac{|x_*-x_*^\prime|}{\epsilon_*}\right)\bigg] e^{(x_*-x_*^\prime)^2/\epsilon_*^2},
\end{equation}
which can then be combined with the portion of the integral with $x_*^{\prime}$ dependence to obtain 
\begin{align}
        &v^{us}_*(x_*,0,t_*)= \nonumber \\
        &\int \limits_0^{t_*} \frac{C_y(s)}{2\pi\epsilon_*^4} \int \limits_{-\infty}^{+\infty} \text{sgn}(x_*-x_*^\prime) (x_*^\prime+s-t_*) \bigg[1-\erf\left(\frac{|x_*-x_*^\prime|}{\epsilon_*}\right)\bigg] e^{(-(x_*^\prime+s-t_*)^2+(x_*-x_*^{\prime})^2)/\epsilon_*^2} \: \mathrm{d}x_*^\prime \mathrm{d} s.
\end{align}
The argument of the exponential function can be rewritten as $-2(x_*+s-t_*)x_*^\prime+x_*^2-(s-t_*)^2$, and with the transformation $x_*^{\prime\prime}=x_*-x_*^{\prime}$ ($\mathrm{d}x_*^{\prime\prime}=-\mathrm{d}x_*^{\prime}$), one obtains
\begin{align}
        v^{us}_*(x_*,0,t_*)= &\int \limits_0^{t_*} \frac{C_y(s)}{2\pi\epsilon_*^4} \int \limits_{+\infty}^{-\infty} (-1)\text{sgn}(x_*^{\prime\prime}) (-x_*^{\prime\prime}+x_*+s-t_*) \bigg[1-\erf\left(\frac{|x_*^{\prime\prime}|}{\epsilon_*}\right)\bigg] \times \nonumber \\
        &e^{(2(x_*+s-t_*)x_*^{\prime\prime}-2(x_*^2+x_*(s-t_*))+(x_*^2-(s-t_*)^2))/\epsilon_*^2} \: \mathrm{d}x_*^{\prime\prime} \mathrm{d} s.
\end{align}
The factor of $(-1)$ can be cancelled by switching the integration limits. Furthermore, the change in sign at $x^{\prime\prime}_*=0$ can be handled by splitting the integral and using the fact that the error function is an odd function ($\erf(-x_*^{\prime\prime})=-\erf(x_*^{\prime\prime})$) in order to eliminate the absolute value in the argument of the error function, which yields
\begin{align}
        v^{us}_*(x_*,0,t_*)= \int \limits_0^{t_*} \frac{C_y(s)}{2\pi\epsilon_*^4}  \int \limits_{0}^{+\infty} & (-x_*^{\prime\prime}+x_*+s-t_*) \bigg[1-\erf\left(\frac{x_*^{\prime\prime}}{\epsilon_*}\right)\bigg] \times \nonumber \\
        &  e^{(2(x_*+s-t_*)x_*^{\prime\prime}-2(x_*^2+x_*(s-t_*))+(x_*^2-(s-t_*)^2))/\epsilon_*^2} \: \mathrm{d}x_*^{\prime\prime} \nonumber \\
         + \int \limits_{-\infty}^{0}& (-1) (-x_*^{\prime\prime}+x_*+s-t_*) \bigg[1+\erf\left(\frac{x_*^{\prime\prime}}{\epsilon_*}\right)\bigg] \times \nonumber  \\
        &  e^{(2(x_*+s-t_*)x_*^{\prime\prime}-2(x_*^2+x_*(s-t_*))+(x_*^2-(s-t_*)^2))/\epsilon_*^2} \: \mathrm{d}x_*^{\prime\prime}  \mathrm{d} s.
\label{eqn:vIndDerivStep}
\end{align}
Subsequent integration in $\mathrm{d}x_*^{\prime\prime}$ yields the solution for the normal induced velocity (equation \ref{eqn:vindUnstSol}).

\subsection{Streamwise induced velocity}\label{appBstreamwise}
The solution for the portion of the integral in equation \ref{eqn:uindUnstBiotSavart} which depends on $y_*^\prime$ is given by
\begin{align}
        - \int \limits_{-\infty}^{+\infty} & \frac{y_*^{\prime2}}{(x_*-x_*^\prime)^2+(-y_*^\prime)^2} e^{-y_*^{\prime2}/\epsilon_*^2} \: \mathrm{d} y_*^\prime = \nonumber \\ &-\sqrt{\pi}\epsilon_* - \pi \, |x_*-x_*^\prime| \, (\erf(|x_*-x_*^\prime|/\epsilon_*)-1) e^{(x_*-x_*^\prime)^2/\epsilon_*^2},
\end{align}
and thus we are left with
\begin{align}
    u^{us}_*&(x_*,0,t_*)= \nonumber \\
    &-\int \limits_0^{t_*} \frac{C_x(s)}{2\pi^2\epsilon_*^4}\int \limits_{-\infty}^{+\infty}  \left(\sqrt{\pi}\epsilon_* + \pi \, |x_*-x_*^\prime| \, (\erf(|x_*-x_*^\prime|/\epsilon_*)-1) e^{(x_*-x_*^\prime)^2/\epsilon_*^2}  \right) e^{-(x_*^\prime+s-t_*)^2/\epsilon_*^2} \mathrm{d}x_*^\prime \mathrm{d} s.
\end{align}
The argument of the exponential simplifies to $-2x_*^\prime (x_*+s-t_*)+x_*^2 - (s-t_*)^2$. The absolute values are handled by splitting the integral in two parts, and we further transform the integrals by substituting $x_*^{\prime\prime}=x_*-x_*^{\prime}$ ($\mathrm{d} x_*^{\prime\prime} = -\mathrm{d} x_*^{\prime}$) to obtain
\begin{align}
    u^{us}_*(x_*,0,t_*)= -\int \limits_0^{t_*} \frac{C_x(s)}{2\pi^2\epsilon_*^4} \int \limits_{0}^{+\infty}  & \left(\sqrt{\pi}\epsilon_* e^{-(x_*-x_*^{\prime\prime}+s-t_*)^2/\epsilon_*^2}\right) + (\pi x_*^{\prime\prime} (\erf(x_*^{\prime\prime}/\epsilon_*)-1) \times \nonumber \\
    & e^{(2x_*^{\prime\prime}(x_*+s-t_*)-2x_*(x_*+s-t_*)+x_*^2-(s-t_*)^2)/\epsilon_*^2}  ) \: \mathrm{d}x_*^{\prime\prime} \mathrm{d} s \nonumber \\
    + \int \limits_{-\infty}^{0} & \left(\sqrt{\pi}\epsilon_* e^{-(x_*-x_*^{\prime\prime}+s-t_*)^2/\epsilon_*^2}\right) + (\pi \, (-1)x_*^{\prime\prime} \, (-\erf(x_*^{\prime\prime}/\epsilon_*)-1) \times \nonumber \\
    & e^{(2x_*^{\prime\prime}(x_*+s-t_*)-2x_*(x_*+s-t_*)+x_*^2-(s-t_*)^2)/\epsilon_*^2}) \: \mathrm{d}x_*^{\prime\prime}\mathrm{d} s,
\end{align}
where we exploited the fact that $\erf(-x_*^{\prime\prime})=-\erf(x_*^{\prime\prime})$. Integration then yields equation \ref{eqn:uindUnstSol}.

\section{Derivation of the vorticity solutions}
\label{appC}
\subsection{Normal forcing}
\label{appCnorm}
Equation \ref{eqn:CyForcingIntegral} can be integrated by parts to obtain
\begin{align}
    \mathcal{I} &=-\int \limits_{x_*-t_*}^{x_*} \frac{\alpha^n_y \cos[\sigma_*^n (\xi-x_*+t_*)] + \beta^n_y \sin[\sigma_*^n (\xi-x_*+t_*)]}{2\pi\epsilon_*^2} \frac{\mathrm{d}}{\mathrm{d} \xi}  \bigg[e^{-(\xi^2+y_*^2)/\epsilon_*^2}\bigg] \mathrm{d} \xi \nonumber \\
    &= \mathcal{I}_1 + \mathcal{I}_2 + \mathcal{I}_3 \nonumber \\
    &= -\frac{\alpha^n_y \cos(\sigma_*^n t_*) + \beta^n_y \sin(\sigma_*^n t_*)}{2\pi\epsilon_*^2}e^{-(x_*^2+y_*^2)/\epsilon_*^2} + \frac{\alpha^n_y}{2\pi\epsilon_*^2} e^{-[(x_*-t_*)^2+y_*^2]/\epsilon_*^2} \nonumber \\
    & + \int \limits_{x_*-t_*}^{x_*} \frac{-\alpha^n_y \sigma_*^n \sin[\sigma_*^n (\xi-x_*+t_*)] + \beta^n_y \sigma_*^n \cos[\sigma_*^n (\xi-x_*+t_*)]}{2\pi\epsilon_*^2} e^{-(\xi^2+y_*^2)/\epsilon_*^2} \mathrm{d} \xi. \label{eqn:C1}
\end{align}
The last integral is computed with the help of the imaginary error function $\erfi(z)=1/i \erf{(iz)}$. To this end, the trigonometric functions are replaced with their representation in terms of complex exponentials
\begin{subequations}
\begin{align}
    \sin[\sigma_*^n(\xi-x_*+t_*)] &= \frac{1}{2i}\bigg[e^{i\sigma_*^n(\xi-x_*+t_*)} - e^{-i\sigma_*^n(\xi-x_*+t_*)}\bigg]\\ 
    \cos[\sigma_*^n(\xi-x_*+t_*)] &= \frac{1}{2}\,\,\bigg[e^{i\sigma_*^n(\xi-x_*+t_*)} + e^{-i\sigma_*^n(\xi-x_*+t_*)}\bigg]
\end{align}
\end{subequations}
which leads to the modified integral
\begin{align}
    \mathcal{I}_3 = \frac{\sigma_*^n}{\epsilon_*^2} e^{-y_*^2/\epsilon_*^2} &\bigg[ (\alpha^n_y i+\beta^n_y) e^{i\sigma_*^n(-x_*+t_*)} \int \limits_{x_*-t_*}^{x_*} e^{i\sigma_*^n \xi - \xi^2/\epsilon_*^2} \mathrm{d} \xi \nonumber \\
    &+  (-\alpha^n_y i+\beta^n_y) e^{-i\sigma_*^n(-x_*+t_*)} \int \limits_{x_*-t_*}^{x_*} e^{-i\sigma_*^n \xi - \xi^2/\epsilon_*^2} \mathrm{d} \xi \bigg].
    \label{eqn:C3}
\end{align}
The two remaining integrals can be solved analytically and are given by
\begin{subequations}
\begin{align}
    \int \limits_{x_*-t_*}^{x_*} e^{i\sigma_*^n \xi - \xi^2/\epsilon_*^2} \mathrm{d} \xi &= -\frac{1}{2}i\sqrt{\pi}\epsilon_* e^{-1/4 (\sigma_*^n)^2 \epsilon_*^2} \bigg[ \erfi\left(\frac{\sigma_*^n\epsilon_*}{2}+\frac{ix_*}{\epsilon_*}\right) - \erfi\left(\frac{\sigma_*^n\epsilon_*}{2}+\frac{i(x_*-t_*)}{\epsilon_*}\right) \bigg], \\ 
    \int \limits_{x_*-t_*}^{x_*} e^{-i\sigma_*^n \xi - \xi^2/\epsilon_*^2} \mathrm{d} \xi &= \frac{1}{2}i\sqrt{\pi}\epsilon_* e^{-1/4 (\sigma_*^n)^2 \epsilon_*^2} \bigg[ \erfi\left(\frac{\sigma_*^n\epsilon_*}{2}-\frac{ix_*}{\epsilon_*}\right) - \erfi\left(\frac{\sigma_*^n\epsilon_*}{2}-\frac{i(x_*-t_*)}{\epsilon_*}\right) \bigg].
\end{align}
\end{subequations}
Thus, the remaining integral $\mathcal{I}_3$ in equation \ref{eqn:C1} is solved, and one obtains the non-zero frequency contributions ($n>0$) of the normal forcing to the vorticity solution given in equation \ref{eq:AnalyticVortSolutionCy}. 

\subsection{Streamwise forcing}
\label{appCstream}
Equation \ref{eqn:CxForcingIntegral} can be rewritten using the exponential expressions for the trigonometric functions as done for the derivation of the normal vorticity, and the resulting integral is given by
\begin{align}
    \omega^{n}_* = -\frac{y_*}{2\pi\epsilon_*^4} e^{-y_*^2/\epsilon_*^2} &\bigg[ (\alpha^n_x - i\beta^n_x) e^{i\sigma_*^n(-x_*+t_*)} \int \limits_{x_*-t_*}^{x_*} e^{i\sigma_*^n \xi - \xi^2/\epsilon_*^2} \mathrm{d} \xi \nonumber \\
    &+  (\alpha^n_x+i\beta^n_x) e^{-i\sigma_*^n(-x_*+t_*)} \int \limits_{x_*-t_*}^{x_*} e^{-i\sigma_*^n \xi - \xi^2/\epsilon_*^2} \mathrm{d} \xi \bigg] .
\end{align}
It is noted that this solution only differs from its normal forcing counterpart (equation
 \ref{eqn:C3}) in terms of the pre-factors of the integrals. The solutions to the integrals themselves are the same as derived in the previous section for the normal forcing. Thus, one obtains the non-zero frequency contributions ($n>0$) of the streamwise forcing to the vorticity solution given in equation \ref{eq:AnalyticVortSolutionCx}. 

\section{A note on velocity sampling in the ALM}
\label{appD}
Both the time and frequency domain solutions of the developed model compute the forcing based on the flow angle determined at the actuator point. This approach was introduced in the foundational ALM work by \citet{sorensen_numerical_2002}. Nevertheless, there is a difference in the definition of the velocity magnitude used for the force calculation between the derived theoretical solutions and the common ALM implementation in LES codes. In principle, the lift and drag coefficients used in step (ii) of the algorithm outlined in Section \ref{sec:TimeDomainSolution} are defined for the free stream velocity $U_\infty$. Since the characteristic velocity scale to non-dimensionalise the equations in this work is chosen to be $U_\infty$, the derived solutions are consistent with this definition. However, common ALM implementations in LES codes not only use the local velocity $\boldsymbol{u}_*^{LES}$ at the actuator point to determine the flow angle for evaluation of the force coefficients but also use its magnitude to directly compute the force, e.g., for the lift force one would have $F^{LES}_L\propto |\boldsymbol{u}_*^{LES}|^2 C_L(\alpha)$. This velocity is affected by the induced velocities such that $|\boldsymbol{u}_*^{LES}|=((1+u^{LES}_*(0,0,t_*))^2+v^{LES}_*(0,0,t_*)^2)^{1/2}$ and thus in case of a non-zero drag coefficient or shed vorticity, it is $|\boldsymbol{u}_*^{LES}|\neq 1$, where it should be noted that for small induced velocities  $|\boldsymbol{u}_*^{LES}|\approx |1 + u^{LES}_*(0,0,t_*)|$. However, the force calculation should be based on the free stream velocity according to  $F_L \propto C_L(\alpha)$. \par 
\cite{martineztossas_optimal_2017} and \cite{caprace_immersed_2020} proposed corrections which are used to determine an estimate of the free stream velocity solely based on the locally sampled velocity at the actuator point. Specifically for this work this would mean that equation \ref{eqn:completeUindSol} for \nolinebreak $u_*$ at the actuator point would be employed to obtain the estimate ${U^{Est}_\infty \approx |\boldsymbol{u}^{LES} - u_*(0,0,t_*)\boldsymbol{i}|}$. In the present work, this issue can be circumvented for the reference LES simulations carried out in Section \ref{Sec:LESvsTheory}, since the free stream velocity $U_\infty$ is known and thus can be directly used for the force calculation. If, on the other hand, the LES the force coefficients would be directly evaluated with the locally sampled velocity magnitude, an additional error between model and LES would be introduced as $F^{LES}_L/F_L= |\boldsymbol{u}^{LES}|^2\approx (1 + u^{LES}_*(0,0,t_*))^2$.

\section{Grid and domain size convergence of the LES setup}\label{appE}
This appendix motivates the chosen LES setup in Section \ref{subsec:LESsetup} by studying the convergence of the angle of attack time history at the actuator point both with grid and domain size. For this purpose we choose a case with a stronger forcing magnitude than any case studied in the validation section (A14-Cxy-S18 for $\epsilon_*=0.25$). The stronger magnitude results in steeper gradients and thus a more conservative setup choice. \par
\begin{figure}
  \centerline{\includegraphics{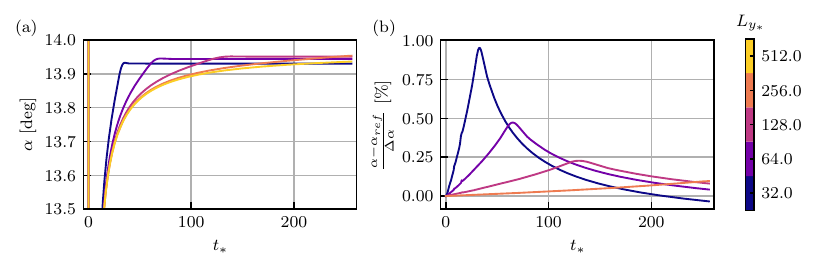}}
  \caption{Domain size convergence study for case A14-Cxy-S18 using $\epsilon_*=0.25$. The simulation time is $T_*=256$, and the resolution of the most inner refinement is $\epsilon_*/\Delta x_*=4$. (a) Angle of attack for the range of semi-domain heights $L_{y_*}=32$ to $L_{y_*}=512$ where the upstream and downstream fetch are scaled accordingly as described in Section \ref{subsec:LESsetup}. (b) Relative error based on the angle of attack step magnitude $\Delta \alpha=18^\circ$. The reference case for the calculation of the error is $L_{y_*}=512$.}
\label{fig:domainSizeConvergence}
\end{figure}

The domain size is completely determined by the simulation time $T_*$ and the length $L_{y_*}$, which is the semi-domain height, but also sets the upstream and downstream fetch. All runs for the convergence study are conducted for $T_*=256$, which is the longest simulation time used for the model validation. The angle of attack time dependence on $L_{y_*}$ is shown in figure \ref{fig:domainSizeConvergence}, where the grid size of the most inner refinement is $\epsilon_*/\Delta x_*=4$ and spans $-4\epsilon_*< x_* < T_*+4\epsilon_*$ and $-4\epsilon_*< y_* < 4\epsilon_*$. It can be seen that the angle of attack evolution is only smooth up to the time instance $t_*$ where the start-up vortex reaches the location $x_*\approx L_{y_*}$ and its induced velocity field, which scales with the inverse of the distance to the vortex core, is altered by the slip boundary conditions applied at the domain top and bottom. Consequently, we choose $L_{y_*}=256$ since we conduct validation LES runs with $T_*=256$. The relative error compared to an even larger domain with $L_{y_*}=512$ (figure \ref{fig:domainSizeConvergence}b) slowly increases with time but does not exceed $0.1\%$, even at the final time instance $t_*=T_*$.\par
\begin{figure}
  \centerline{\includegraphics{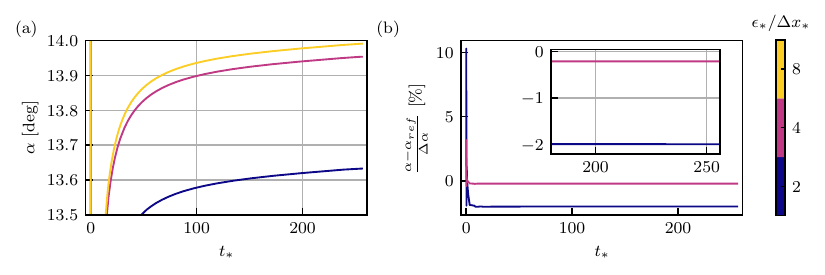}}
  \caption{Grid size convergence study for case A14-Cxy-S18 using $\epsilon_*=0.25$. The simulation time is $T_*=256$ and the semi-domain height is $L_{y_*}=256$. (a) Angle of attack for the range of grid resolutions $\epsilon_*/\Delta x_*=2$ to $\epsilon_*/\Delta x_*=8$ (for the most inner refinement). The semi-domain height for all three cases is $L_{y_*}=256$. (b) Relative error based on the angle of attack step magnitude $\Delta \alpha=18^\circ$. The reference case for the calculation of the error is $\epsilon_*/\Delta x_*=8$.}
\label{fig:gridSizeConvergence}
\end{figure}

Furthermore, the resolution of the most inner refinement is varied between $\epsilon_*/\Delta x_*=2$, $\epsilon_*/\Delta x_*=4$ and $\epsilon_*/\Delta x_*=8$ using the chosen $L_{y_*}=256$. Note that the resolution of the base grid is constant, and the resolution is varied by adding additional refinement levels around the actuator point and the wake region. The normal extent of the most inner refinement is always $-4\epsilon_*< y_* < 4\epsilon_*$. The streamwise extent of the two coarser resolutions is $-4\epsilon_*< x_* < T_*+4\epsilon_*$, whereas the highest resolution case only adds an additional refinement level between $-4\epsilon_*< x_* < 16+4\epsilon_*$. The evolution of the angle of attack on these three different grid resolutions is shown in figure \ref{fig:gridSizeConvergence}. Contrary to the domain size convergence, the relative angle of attack error for the grid resolution convergence stays constant after the start-up phase. This is the case because it is governed by the strength of the start-up vortex, which is determined during the initial phase of the step response. The relative error for the $\epsilon_*/\Delta x_*=4$ case is below $0.25\%$ after the first few time steps (figure \ref{fig:gridSizeConvergence}b). We conservatively choose here a resolution of $\epsilon_*/\Delta x_*=8$ for all model validation runs, which ensures fully converged LES results.

\FloatBarrier

\bibliographystyle{jfm}
\bibliography{jfm}

\end{document}